\documentstyle[aas2pp4]{article}

\lefthead{Meier}
\righthead{Multi-Dimensional Astrophysical Structural and Dynamical Analysis}

\def\inode{{\hat{\imath}}}
\def\jnode{{\hat{\jmath}}}
\def\knode{{\hat{\it{k}}}}
\def\lnode{{\hat{\ell}}}
\def\bm#1{\mbox{\boldmath{$#1$}}}
\def\bs#1{\mbox{\boldmath{$#1$}}}
\def\ilocalnode{{\tilde{\imath}}}

\received{ }

\slugcomment{Accepted for publication in The Astrophysical Journal.}

\begin{document}

\hyphenation{ }

\title{Multi-Dimensional \\
    Astrophysical Structural and Dynamical Analysis \\
    I.  Development of a Nonlinear Finite Element Approach}

\author{D. L. Meier}
\affil{Jet Propulsion Laboratory, California Institute of Technology,
    Pasadena, CA 91109}



\begin{abstract}
A new field of numerical astrophysics is introduced which addresses the 
solution of large, multidimensional structural or slowly-evolving problems 
(rotating stars, interacting binaries, thick advective accretion disks, four 
dimensional spacetimes, {\it etc.}), {\em as well as} rapidly-evolving systems.  
The technique employed is the Finite Element 
Method (FEM), which has been used to solve engineering structural problems for more 
than three decades.  The approach developed herein has the following key features:
\begin{enumerate}
\item The computational mesh can extend into the time dimension, as well as space 
--- generally only a few cells deep for most (flat-space) astrophysical 
problems, but throughout spacetime for solving Einstein's field 
equations.
\item When time is treated as a mesh dimension, virtually all equations 
describing the astrophysics of continuous media, including the field equations, 
can be written in a compact form 
similar to that routinely solved by most engineering finite element codes 
(albeit for nonlinear equations in a four-dimensional spacetime instead of 
linear ones in two or three space dimensions):  the divergence of a generalized 
stress tensor equals a generalized body force vector, both of which are functions 
only of position, the state variables and their gradients.  
\item The transformations that occur naturally in the four-dimensional FEM 
possess both coordinate and boost features, such that
\begin{enumerate}
\item although the computational mesh may have a complex, non-analytic, 
curvilinear structure, and may be adapted to the geometry of the problem, the 
physical equations still can be written in a simple coordinate system that is 
independent of the mesh structure.
\item if the mesh has a complex flow velocity with respect to coordinate space, 
the transformations will form the proper advective derivatives, automatically 
converting the equations to arbitrary Lagrangian-Eulerian.
\end{enumerate}
\item Only relatively simple {\em differential} equations need to be encoded.  The 
complex {\em difference} equations on the arbitrary curvilinear grid are generated 
automatically by the FEM integrals.  A different integration method must be used 
for equations of odd and even order.
\end{enumerate}

This first paper concentrates on developing a robust and widely-applicable set of 
techniques using the nonlinear FEM and presents some examples.  The second 
paper in this series will deal with making the method fast and efficient, so that 
large, astrophysically-interesting computational meshes can be employed.
\end{abstract}


%
%


\section{Introduction}

The first problem to be solved with the techniques of numerical astrophysics 
was the structure and evolution of stars --- an ``implicit'' problem that 
involves a static or slowly-evolving structure (\cite{chandra57};
\cite{am65}).  Its solution consists 
of determining the state variables of the fluid (density, temperature, 
pressure, flux of radiation, composition) at each radius in the stellar 
interior, and is obtained by relaxing a large set of coupled nonlinear 
difference equations (derived from the differential equations  of stellar 
structure) along with boundary conditions at the stellar center and surface.  
Radial stellar structure is only a one-dimensional problem and, 
while once considered difficult and CPU-intensive, now is solved easily on 
personal computers (PCs).  Since then, many other fields of numerical 
astrophysics have been developed:  ``explicit'' hydrodynamic simulations of 
explosive and jet phenomena (\cite{norman97}); N-body and smooth particle 
hydrodynamics (SPH) of discrete or semi-discrete systems of particles 
(\cite{dub97}; \cite{monaghan92}); Monte Carlo simulations of radiation flow 
(\cite{leahy97}; \cite{ph98}); {\it etc}.  All have matured to the point where 
the solution of three-dimensional, time-dependent problems is not uncommon. 

Ironically, however, only modest progress has been made in extending the 
original implicit problems into several dimensions:  rapidly rotating 
stars, evolving and interacting binaries, detailed accretion disk structure and 
evolution, etc.  There are several important reasons for this.  Firstly, the 
geometry of these systems is unknown until the problem is solved.  For example, 
the shape of the outer surface of a rotating star or (possibly thick and 
advective) accretion disk will be part of the solution, and the shape of a 
rapidly rotating stellar core may have a different oblateness (or even 
prolateness) from that of the 
outer envelope.  No numerical method capable of operating in nearly-arbitrary 
geometries has been applied extensively in astrophysics.  Instead, one either 
has assumed spherical symmetry and treated only slowly-rotating, perturbation 
problems (\cite{kt70}), or has assumed that the isosurfaces of the state 
variables are coincident (which implies rotation on cylinders via von Zeipel's 
theorem) and again solved an essentially one-dimensional, or limited 
two-dimensional problem (\cite{em91}; \cite{clement94}).
Recently, some progress for two- and three-dimensional stellar models has been 
made using a multi-domain approach and treating the stellar surface as a 
discontinuity (\cite{bgm98}).

Secondly, even if a general geometrical method applicable to large 
numbers of two-dimensional and three-dimensional problems could be developed, 
the current stellar 
structure methods for solving the immense system of simultaneous nonlinear 
equations would take a prohibitively long amount of CPU time and memory.  
For example, a relatively modest problem with $256^3$ grid points and 
ten state variables at each point would generate a banded matrix $10^{8.2} 
\times 10^{8.2}$ in size, taking up at least $10^{14.9}$ bytes (0.9 PB) for the 
non-zero elements. Direct inversion techniques, similar to the Henyey 
method commonly used in stellar structure (which take the bandedness 
into account), would take about a thousand years to invert this matrix once 
on a large parallel supercomputer like the Cray T3D, with perhaps $10^4$ or 
more such inversions necessary for a complete stellar evolution 
model\footnote{For this calculation it is assumed that the storage is 
proportional to the bandwidth $B~(\sim 10^{5.8})$ of the matrix times its 
length L and the time to invert is proportional to $B^2 \, L$.}.  

Fortunately, there exist techniques for solving both of these problems that are 
well developed and have been in use in the engineering field for many years 
(although it is still rare to see both used at the same time).  The Finite 
Element Method (FEM), introduced more than four decades ago and 
the preferred method of treating multidimensional structural engineering 
problems since the late 1960s (\cite{zienk}), approximates objects as 
distorted lattices of small structural members called elements.  For solving 
large systems of coupled 
equations generated by such grid problems, the multigrid method was introduced 
in the 1970s (\cite{brandt77}) and is now beginning to be used in astrophysics 
as well (\cite{tru98}; \cite{norman98}).  This approach dispenses with the 
large matrix, cleverly reaching a solution after a few sweeps of the mesh.   
Together, these techniques promise to make multidimensional astrophysical 
structural problems possible, and bring the time to solve them within an order 
of magnitude or so of that for explicit problems.  When coupled with the 
continued expected increase in speed of computers over the next few decades 
(which has averaged about a factor of 2 every 2 years for the past 20 years), 
it is not inconceivable that three-dimensional {\em structure} problems will 
soon be solved routinely on future models of PCs and that four-dimensional 
problems will become commonplace on supercomputers.

This first paper deals with the development of the general geometrical method 
for solving multi-dimensional structure and evolution problems.  At this stage the 
speed and efficiency of execution of the method will not be a concern; the focus 
will be only on producing a robust and widely-applicable set of useful techniques.  
Our goal will be to determine the essential features of most astrophysical systems 
of equations and the geometrical demands they place on the numerical method.  These 
properties will then be encoded at the outset, ensuring some measure of generality.
The second section describes the set of equations that can be addressed with 
nonlinear astrophysical finite element analysis and develops a method for solution on 
fixed (non-moving) grids.  In section 3, this then is generalized to include situations 
where the positions of the grid points are part of the solution and the grid can change 
with time.  Finally, tests and examples are given, using the author's code, including 
rotating polytropic star models.

\section{The Basic Four-Dimensional, Nonlinear Finite Element Method on Fixed Grids}

This section describes the techniques used in the author's computer code, 
entitled GENRAL, for solving general astrophysical problems.  
It utilizes the techniques of finite element analysis (FEA) --- in use in the 
field of engineering for some time --- but generalizes them to nonlinear 
equations in four dimensions, instead of linear equations in two or three 
dimensions.  
For this initial development, it is assumed that the coordinates of the grid 
points (or ``nodes'') at which the variables are evaluated do not change while 
the solution is being computed.

\subsection{General Form for Equations of Continuum Astrophysics}

Appendix A shows that the differential equations of continuum 
astrophysics in curved spacetime can be cast into the generic form
\begin{eqnarray}
\Re_{q} & \equiv & 
( T^{\beta}_{q}({\bm{\nabla w}},~{\bm{w}},~{\bm{x}}) \, \sqrt{-g} )_{, \beta} 
\nonumber \\
& & \, - \, F_{q}({\bm{\nabla w}},~{\bm{w}},~{\bm{x}}) \, \sqrt{-g} 
\nonumber \\
\label {general_eq}
& = & 0 
\end{eqnarray}
where $[{\bm{w}}]^v = w^v$ is a generalized solution vector holding all of the 
$v = 1, ..., V$ unknown state variables; 
$\Re_{q}$ is a generalized residual for each of the $q = 1, ..., V$ equations 
(which will be forced to zero through numerical relaxation techniques); 
$T_{q}$ and $F_{q}$ are, respectively, the generalized stress tensor and 
force vector for these equations; and $g$ is the determinant of the metric tensor ${\bm{g}}$ 
([--+++]-signature).\footnote{Throughout the paper the notation of \cite{mtw} is 
used, with Greek letters 
indicating coordinate indices in four-dimensional spacetime ($\alpha = 0, 1, 2, 3$) 
and Latin ``integer'' letters indicating three-space indices only ($i=1,2,3$).  
The comma denotes ordinary differentiation with respect to the coordinates, while a 
semicolon will denote covariant differentiation.  Repeated indices indicate summation over 
the entire range of those indices (the Einstein summation convention), so that
\begin{displaymath}
g^{\alpha \mu} g_{\mu \beta , \gamma} \; \equiv \; \sum_{\mu = 0}^3 \, 
g^{\alpha \mu} \frac{\partial g_{\mu \beta}}{\partial x^{\gamma}} 
\end{displaymath}
A raised index indicates contravariant properties of the tensor and a lowered index 
indicates covariant properties.  Note that $w^v$ is written 
as a contravariant vector with a raised index, like the coordinates $x^{\alpha}$; this is 
partly for convenience (to facilitate the summation convention) and partly to 
draw attention to the variables as generalized coordinates of the system.}
In a similar manner, the boundary conditions on these equations can be cast as
\begin{eqnarray}
\Re_{r} & \equiv & 
[ \, t^{\beta}_{r}({\bm{\nabla w}},~{\bm{w}},~{\bm{x}}) \, \sqrt{-g} \, ]_{, \mu} 
{{\cal{S}}^{\mu}}_{\beta}
\nonumber \\
& & \, + \, [ \, h_{r}({\bm{S \cdot \nabla w}},~{\bm{w}},~{\bm{x}}) \, \sqrt{-g} \, ]_{, \mu} 
{{n}^{\mu}}
\nonumber \\
& & \, - \, f_{r}({\bm{\nabla w}},~{\bm{w}},~{\bm{x}}) \, \sqrt{-g} 
\nonumber \\
\label {general_bc}
& = & 0 
\end{eqnarray}
for each of the $r = 1, ..., R$ boundary conditions.  In general, $R \neq V$ 
since, depending on the highest order derivative in $\Re_{r}$, there may be 0, 
1, or 2 associated boundary conditions. 
The last term in (\ref{general_bc}) is adequate for handling Dirichlet, 
Neumann, and mixed boundary conditions, such as the radiative condition at a 
stellar or accretion disk surface.  The first two terms are necessary for 
including constraints on field equations.  
{\bm{\cal{S}}} is the projection tensor along the boundary $\partial \Omega$ 
and orthogonal to the boundary normal ${\bm{n}}$ 
\begin{equation}
\label {projection_tensor}
{\bm{\cal{S}}} \; \equiv \; {\bm{n \otimes n}} \, + \, {\bm{g}}
\end{equation}
where ${\bm{n \cdot n}} = -1$, ${\bm{n \cdot \cal{S}}} = 0$, and 
${\bm{\cal{S} \cdot \cal{S}}}  = {\bm{\cal{S}}}$, 
``$\bm{\otimes}$'' is the outer (dyadic) product, 
and ``$\bm{\cdot}$'' is the inner (scalar) product.  
{\bm{\cal{S}}} causes the divergence in equation (\ref{general_bc}) to be 
performed on the boundary only and the derivative normal to the boundary to be 
only first order in $n^{\mu} \, \partial/\partial x^{\mu}$.  An alternative 
form for (\ref{general_bc}) is 
\begin{eqnarray*}
\Re_{r} & \equiv & 
[ \, {t'}^{\beta}_{r}({\bm{\nabla w}},~{\bm{w}},~{\bm{x}}) \, \sqrt{-g} \, ]_{, \mu} 
{{\cal{S}}^{\mu}}_{\beta}
\\
& & \, + \, [ \, {h'}^{\mu}_{r}({\bm{S \cdot \nabla w}},~{\bm{w}},~{\bm{x}}) \, \sqrt{-g} \, ]_{, \mu} 
\\
& & \, - \, {f'}_{r}({\bm{\nabla w}},~{\bm{w}},~{\bm{x}}) \, \sqrt{-g} 
\\
& = & 0 
\end{eqnarray*}
which also has no second derivatives in $n^{\mu} \, \partial/\partial x^{\mu}$.  

It is important to note that equation (\ref{general_eq}) includes not only 
structural and steady problems, but also evolving ones as well.  
For these cases, in addition to having three spatial coordinates, the 
computational grid can extend into the fourth (time) dimension, possibly from 
the initial time step or hypersurface to the final one.  
While certainly increasing the computational and memory load on the computer, 
this approach will have distinct advantages over conventional 
approaches to initial-value problems.  

The boundary conditions are not necessarily completely described by equation 
(\ref{general_bc}).  While it is well known that, {\em physically}, the 
boundary of a boundary is zero ($\partial \partial \Omega = 0$), 
computationally one often introduces sub-boundaries by truncating the mesh or 
imposing symmetries on the problem.  These conditions produce right-angle kinks 
in the boundary, where ${\bm{n}}$ suddenly rotates by $90^{\circ}$ and the 
boundary conditions abruptly change.  Such boundary corners occur, {\it e.g.}, 
where the $t = 0$ initial hypersurface intersects the world line of a stellar 
surface or (in the case of axisymmetry or plane symmetry) where the stellar 
surface intersects the symmetry axis or plane.  For example, when solving Maxwell's 
evolutionary equations on the four-dimensional domain $\Omega$ under such 
conditions, they will be bounded on the $t = 0$ portion of $\partial \Omega$ 
by the initial value (solenoidal and Coulomb) constraints; these will be 
bounded further at an external stellar 2-surface $\partial \partial \Omega$; 
and these may be bounded still further by the rotation axis or equatorial 
plane at edges $\partial \partial \partial \Omega$.  Therefore, additional 
equations, similar to (\ref{general_bc}), with successive projection of the 
first two terms into the sub-boundaries of lower dimension, may be needed 
until one reaches the zero-dimensional $\partial \partial \partial \partial 
\Omega$ (endpoints of line segments) where the conditions become simply 
\begin{equation}
\label {simple_bc}
\Re_{r} \; \equiv \; - \, f_{r}({\bm{\nabla w}},~{\bm{w}},~{\bm{x}}) \; = \; 0
\end{equation}
In the examples in this paper all boundary conditions are of the simple form 
(\ref{simple_bc}), but in general astrophysical situations the form 
(\ref{general_bc}) will be needed.

\subsection{Continuous Solution:  The Element Mesh}

Formally, in the finite element method (FEM) the computational domain $\Omega$ is 
subdivided not into nodes, but into sub-domains ($\delta \Omega$) called ``elements'' --- similar to 
``zones'' or ``cells'' in the finite difference method (FDM).  (Nodes come later, and then 
only to facilitate the element process.)  The elements are constructed in 
such a way that each function $w^v$ is continuous over the entire domain, but 
its derivatives are only piecewise continuous; {\it i.e.}, $w^v$ is 
continuously differentiable only within each element.
The $x^{\alpha}$ are treated in the same manner; they also are continuous across element 
boundaries with no spatial ``gaps'' between elements.  

Although a variety of generic element shapes can be used, the most 
common are triangular and quadrangular.  Of course, these assume higher-order 
shapes in three and four dimensions ({\it i.e.}, equilateral triangles, 
tetrahedra, simplices [hyper-tetrahedra]; squares, cubes, and hypercubes), but 
they still shall be referred to here as the triangular and quadrangular classes.  
In GENRAL, elements of the quadrangular type are used exclusively because of 
their convenience. 
The computational domain is filled with a topologically rectangular set of 
\begin{equation}
{\cal{E}} \; = \; \prod_{\alpha ' = 0}^{D-1} {\aleph}_{\alpha '}  
\end{equation}
of these building blocks, where $D (\leq 4)$ is the dimensionality of the 
problem, and $\aleph_{\alpha '}$ is the number of elements along each mesh 
dimension $\alpha '$.  Each element that borders the domain has one surface 
lying on the boundary that itself is an element of dimension $D-1$.  The total 
number of such boundary elements enclosing this rectangular mesh is a sum over 
the rectangular faces
\begin{equation}
{\cal{B}} \; = \; 2 \sum_{\alpha ' = 0}^{D-1} { ~~ \prod_{\beta ' \neq \alpha '}^{D-1} 
{\aleph_{\beta '} } } 
\end{equation}

The element mesh can be distorted by stretching, compressing, bending, or even twisting 
it to conform to the geometry of the domain (as, for example, in a curvilinear coordinate 
system). 
In the engineering FEM this coordinate transformation is called the ``isoparametric'' 
transformation, because coordinate values $x^{\alpha}$ and the variables $w^v$ are 
specified at the same nodal points.  In general relativity this transformation 
is the generalized Lorentz transform
\begin{equation}
\label{isoparm_general}
{{\cal{L}}^{\alpha}}_{\alpha '} \; \equiv \; \frac{\partial x^{\alpha}}{\partial \xi^{\alpha '}}
\end{equation}
with inverse
\begin{equation}
\label{isoparm_general_inv}
{{\cal{L}}^{\alpha '}}_{\alpha} \; \equiv \; \frac{\partial \xi^{\alpha '}}{\partial x^{\alpha}}
\end{equation}
where $\xi^{\alpha '}$ is the coordinate in mesh space, with range 
$0 \leq \xi^{\alpha '} \leq 1$ in each dimension $\alpha '$.  Basis vectors 
along the mesh coordinate direction $\alpha '$, and corresponding 1-forms, are
\begin{displaymath}
{[{\bm{e}}_{\alpha '}]^{\alpha}} \; = \; {{\cal{L}}^{\alpha}}_{\alpha '} ~~~~~
{[{\bm{\omega}}^{\alpha '}]_{\alpha}} \; = \; {{\cal{L}}^{\alpha '}}_{\alpha} 
\end{displaymath}
Appendix B discusses conditions that may need to be 
satisfied by this transformation.  However, unless one wishes to use the mesh 
as an actual Lorentz frame of reference, or wants to follow the evolution of 
all wave phenomena, only the Jacobi condition is necessary for numerical 
stability
\begin{equation}
\label{jacobi_cond}
{\cal{L}} \; \equiv \; {\rm det} ||{\bm{\cal{L}}}|| \neq 0
\end{equation}

\subsection{The Choice of a Basic Coordinate System}

In the past, when developing a finite difference numerical simulation 
code, for example, it has been customary 
(and considered necessary) to write the differential equations in the same 
coordinate system described by the computational mesh.  That is, if the mesh is 
spherical-polar, then the equations are written in spherical-polar coordinates, 
and so on.  However, in the numerical method developed in this paper, this 
degeneracy is neither necessary nor desirable, as the {\em mesh} coordinate system 
is unknown until the problem is solved.  

To allow for an arbitrary, unknown mesh, the differential equations will be written 
in a ``basic'' or ``real-space'' ($x^{\alpha}$) system which does not change as 
the calculation proceeds.  The derivatives still will be computed in the mesh 
($\xi^{\alpha '}$) system, but, in order to use them in the differential 
equations, will then be transformed to the basic system
\begin{equation}
{[{\bm{\nabla w}}]^v}_{\alpha} \; \equiv \; {w^v}_{, \alpha} \; = \; 
{{\cal{L}}^{\alpha '}}_{\alpha} {w^v}_{, \alpha '} 
\end{equation}
using the isoparametric/Lorentz transformation.  

The choice of coordinate system for the mesh is determined by how one 
lays out the elements in real space.  That is, 
\begin{equation}
g_{\alpha ' \beta '} \; = \; {{\cal{L}}^{\alpha}}_{\alpha '} \, 
{{\cal{L}}^{\beta}}_{\beta '} \, g_{\alpha \beta} 
\end{equation}
gives the metric coefficients in mesh space.  However, one still needs to choose 
a system in which to write the differential equations;  but, 
since the computer will be doing all the curvilinear work for us, one 
can select a very simple basic system, keeping the coordinates
as Cartesian (or as Minkowskian) as possible.  For example, in axisymmetric
problems, cylindrical coordinates will be used, not spherical-polar.  
For three-dimensional problems, Cartesian coordinates will be used, no matter 
how spherical the star or flattened the accretion disk.
Any curvilinear properties of the metric {\em orthogonal} to the computational domain 
will be embodied in the volume element $\sqrt{-g}$, and curvilinear behavior 
{\em within} the domain will be handled by the isoparametric transformation.

\subsection{Discrete Solution:  The Nodal Mesh and Interpolation Scheme}

As with all continuum numerical methods, the solution is expressed as a finite 
set of discrete values.  In the FDM these are values of the solution at specified 
points (nodes) in space;  in spectral methods these 
are coefficients of basis or interpolation functions.  In the FEM, these 
discrete values are {\em both} nodal values {\em and} basis function coefficients.
That is, {\em the FEM has properties of both finite difference and spectral methods}.

The finite element nodes are distributed within each element in such a way that the 
$w^v$ can be interpolated across the element in each dimension with at least linear 
accuracy or better.  
For quadrangular elements, the simplest approach is 
to fill each element box with a (possibly hyper-) cubic mesh of $(\wp_{\alpha '} + 1)$ nodes 
per dimension $\alpha '$, where $\wp_{\alpha '}$ is the order of interpolation 
in that dimension, and nodes are shared by adjacent elements at all the interfaces 
(corners, edges, faces, and hyperfaces).  For a problem of total number of 
dimensions $D$, the total number of nodes describing each {\em element} is, then, 
\begin{equation}
I \; = \; \prod_{\alpha ' = 0}^{D-1} (\wp_{\alpha '} + 1)
\end{equation}
That is, for four-dimensional elements, $I=16$ for first-order (linear) interpolation, 
$I=81$ for second-order (quadratic) interpolation, and $I=256$ for third-order (cubic) 
interpolation --- just within each element.  The total number of nodes in the entire mesh is
\begin{equation}
{\cal{I}} \; = \; \prod_{\alpha ' = 0}^{D-1} ({\aleph}_{\alpha '} \wp_{\alpha '} + 1)
\end{equation}
(no sum on $\alpha '$).  
The number of nodes {\em on each element's boundary} is the total minus those 
in the interior
\begin{equation}
K \; = \; I \, - \, \prod_{\alpha ' = 0}^{D-1} (\wp_{\alpha '} - 1)
\end{equation}
and for the entire mesh
\begin{equation}
{\cal{K}} \; = \; {\cal{I}} \, - \, \prod_{\alpha ' = 0}^{D-1} ({\aleph}_{\alpha '} 
\wp_{\alpha '} - 1)
\end{equation}

No matter what the value of $\wp_{\alpha '}$, in these simple cases the basis 
functions in mesh space for a node $\inode$ within a given element $e$ are 
products of Lagrange interpolation polynomials ${\pounds}_{\inode e \alpha '}$ 
in each dimension $\alpha '$ 
\begin{eqnarray}
\label {lagrange_shapes}
{{\cal{N}} '}_{\inode e}({\bm{\xi}}) & = \; & {\pounds}_{\inode e}({\bm{\xi}}) \nonumber 
\\
                                & \equiv \; & \prod_{\alpha ' = 0}^{D-1} 
                                  \, {\pounds}_{\inode e \alpha '}({\xi}^{\alpha '}) \nonumber 
\\
                                & = \; & \prod_{\alpha ' = 0}^{D-1} 
                                  \, \prod_{\jnode \neq \inode}^{I} 
                          \, \frac{(\xi^{\alpha '}            - {\xi}^{\alpha '}_{\jnode e})}
                                  {(\xi^{\alpha '}_{\inode e} - {\xi}^{\alpha '}_{\jnode e})}
\end{eqnarray}
where $\xi^{\alpha '}_{\inode e}$ is the mesh coordinate value at node $\inode$ in element $e$, 
${{\cal{N}} '}_{\inode e}$ is the contribution from that element to the basis or ``shape'' 
function for node $\inode$, as measured
in the mesh (primed) system. The range of the nodal indices in the entire mesh is $\inode, 
\jnode = 1, ..., {\cal{I}}$, but the product in equation 
(\ref{lagrange_shapes}) only runs over the nodes within element. The total 
basis function for node $\inode$ is, then, a sum over the element 
contributions 
\begin{equation}
{{\cal{N}} '}_{\inode}({\bm{\xi}}) \; = \; \sum_{e} {{\cal{N}} '}_{\inode e}({\bm{\xi}}) 
\end{equation}
(which actually involves only those elements containing that node). 
Note that each shape function attains unit value at its own node and zero at 
all other nodes in its associated elements (and in the mesh as well)
\begin{equation}
{{\cal{N}} '}_{\inode}({\bm{\xi}}_{\jnode}) \; = \; \delta_{\inode \jnode}
\end{equation}

Often the body-centered nodes, and sometimes even face-centered nodes, are  
removed from the standard Lagrangian elements elements to form the 
so-called ``serendipitous'' elements (\cite{zienk}).  In that case, if node $\lnode$ 
is removed, then the basis functions are given by the normal Lagrange shape function 
with the Lagrange shape function for that missing node subtracted off 
\begin{equation}
\label {serendip_shapes}
{{\cal{N}} '}_{\inode}({\bm{\xi}}) \; = \; {\pounds}_{\inode}({\bm{\xi}}) \, - 
{\pounds} _{\lnode}({\bm{\xi}})
\end{equation}
with $\lnode \neq \inode$.  

Most of the properties of Lagrangian elements can be illustrated in one 
dimension.  Figure \ref{shapes_fig} shows a simple 1-dimensional, 5-node mesh and 
its discretization in linear and quadratic elements.  Note that 
interior shape functions have continuous derivatives at their respective nodes, 
while shape functions on 
element boundaries have discontinuous derivatives.  (The latter also involve more 
nodes as they are composed of shape function pieces from adjacent 
elements.)  The FEM, therefore, can be considered to be a multi-domain 
spectral method with each of the thousands to millions of elements being a 
separate domain. 

Because the shape functions are continuous throughout $\Omega$, the solution 
${\bm{w}}$ and the coordinates ${\bm{x}}$ are truly continuous functions of 
position in the mesh: 
\begin{eqnarray}
\label {w_fns_mesh_space}
w^v ({\bm{\xi}})        & = & {{\cal{N}} '}_{\inode} ({\bm{\xi}}) \, w^v_{\inode} 
\\
x^{\alpha} ({\bm{\xi}}) & = & {{\cal{N}} '}_{\inode} ({\bm{\xi}}) \, x^{\alpha}_{\inode} 
\end{eqnarray}
like spectral methods but unlike the FDM where interpolation is only an 
{\it ad hoc} addition to the scheme.
Also, because a unique inverse relation ${\bm{\xi}} = {\bm{\xi}}({\bm{x}})$ 
exists, the variables have an implicit function of position in real space 
\begin{equation}
w^v ({\bm{x}}) \; = \; w^v ({\bm{\xi}}({\bm{x}}))
\end{equation}

\subsection{Formation of Derivatives and the Differential Equations}

We now have a numerical procedure for computing the derivatives of the $w^v$ with 
respect to $x^{\alpha}$.  First, the 
coordinate transformation matrix is formed
\begin{equation}
{{\cal{L}}^{\alpha}}_{\alpha '} \; = 
\; {{\cal{N}} '}_{\inode , \alpha '} \, x^{\alpha}_{\inode} 
\end{equation}
and then inverted to obtain ${{\cal{L}}^{\alpha '}}_{\alpha}$.  Then the derivatives 
of the variables are computed in mesh space
\begin{equation}
{w^v}_{, \alpha '} \; = \; {{\cal{N}} '}_{\inode , \alpha '} w^v_{\inode}
\end{equation}
and, finally, transformed to real space by the chain rule
\begin{equation}
\label {w_ders_mesh_space}
{w^v}_{, \alpha} \; = \; {{\cal{L}}^{\alpha '}}_{\alpha} \, 
{{\cal{N}} '}_{\inode , \alpha '} \, w^v_{\inode} 
\end{equation}

Although not usually used in practice in the actual computer code, it is sometimes 
useful for analytic purposes to express the shape functions in real space coordinates 
and use them to interpolate the $w^v$ and compute their derivatives
\begin{eqnarray}
\label{w_fns_real_space}
w^v ({\bm{x}})           & = & {{\cal{N}}}_{\inode} ({\bm{x}}) \, w^v_{\inode} 
\\
\label{w_ders_real_space}
{w^v}_{, \alpha} ({\bm{x}}) & = & {{\cal{N}}}_{\inode , \alpha} ({\bm{x}}) w^v_{\inode}
\end{eqnarray}
The real-space ${{\cal{N}}}_{\inode}$ also have the normalized property at their respective 
nodes
\begin{displaymath}
{{\cal{N}}}_{\inode}({\bm{x}}_{\jnode}) \; = \; \delta_{\inode \jnode}
\end{displaymath}
By comparing (\ref{w_fns_real_space}) and (\ref{w_ders_real_space}) with 
with (\ref{w_fns_mesh_space}) and (\ref{w_ders_mesh_space}), one concludes 
that the real-space basis functions and their derivatives are
\begin{eqnarray}
{{\cal{N}}}_{\inode} ({\bm{x}})         & = \; & {{\cal{N}} '}_{\inode} ({\bm{\xi}}({\bm{x}})) 
\\
{{\cal{N}}}_{\inode , \alpha} ({\bm{x}}) & = \; & {{\cal{L}}^{\alpha '}}_{\alpha}  \, 
                                       {{\cal{N}} '}_{\inode , \alpha '} ({\bm{\xi}}({\bm{x}})) 
\end{eqnarray}

With expressions for the $w^v$ and ${w^v}_{, \alpha}$ (either in mesh or real 
space) we now can calculate the residuals $\Re_{q}$ (equation \ref{general_eq}) 
at {\em any} point in the domain $\Omega$ and $\Re_{r}$ (equation 
\ref{simple_bc}) at {\em any} point on the boundary $\partial \Omega$, not 
just at the nodes.

\subsection{Generation of the Nodal (``Difference'') Equations:  The Weighted 
Residual Method}

\subsubsection{General Formulation}

The next step in the development of the astrophysical FEM is to construct a 
set of $V {\cal{I}}$ equations for the $V {\cal{I}}$ shape function 
coefficients ($w^v_{\inode}$) that fully describe $w^v ({\bm{x}})$.  This is 
accomplished by integrating the physical differential equations 
(\ref{general_eq}) and/or boundary conditions (\ref{simple_bc}) over a 
function ${\cal{W}}_{\inode} ({\bm{x}})$ which peaks near (but not necessarily 
at) node $\inode$ and falls to zero far from that node.  
This produces $V{\cal{I}}$ discrete nodal equations 
\begin{equation}
\label {strong_form}
\Im_{q \inode} (w^v_{\jnode}) \; = \; \int_{\Omega} \, {\cal{W}}_{\inode} ({\bm{x}}) 
\, \Re_{q} (w^v_{\jnode}) \, d \Omega \; = \; 0
\end{equation}
Relaxation schemes in the code then attempt to force the nonlinear 
$\Im_{q \inode}$ to zero.  In principle, each $\Im_{q \inode}$ is a function of 
{\em all} of the $w^v_{\jnode}$.  However, in practice, because 
${\cal{W}}_{\inode}$ is peaked near node $\inode$, $\Im_{q \inode}$ involves 
only nodes local to $\inode$ --- in fact, only nodes in those elements 
containing node $\inode$.  The $\Im_{q \inode}$, therefore, are more similar 
to difference equations than to spectral equations and, when linearized, 
produce a banded rather than filled matrix.

Because $\Re_{q}$ can contain second derivatives, the integral in equation 
(\ref{strong_form}) cannot be performed uniquely for every node using only the 
interpolation within a single element.\footnote{The problem occurs at boundary nodes, 
where one needs information in adjacent elements.  For example, in one dimension 
linear shape functions have zero second derivative, so they cannot represent the kernel 
at all.  Quadratic shape functions do have a second derivative, but only one unique value 
within a given element, which is valid for the central node, but not for the boundary nodes.}  
The standard solution to this problem, and the key step in the finite element process,
is to integrate the weighted residual by parts to arrive at the so-called ``weak'' form
\begin{eqnarray}
\label {weak_form}
\Im_{q \inode} (w^v_{\jnode}) & = & \int_{\partial \Omega} \, {\cal{W}}_{\inode} ({\bm{x}}) 
\, T^{\beta}_{q} \, d(\partial \Omega)_{\beta} - \nonumber 
\\
& & \int_{\Omega} \, ( {\cal{W}}_{\inode} ({\bm{x}})_{, \beta} 
T^{\beta}_{q} \, + \, {\cal{W}}_{\inode} ({\bm{x}}) \, F_{q} ) \, d \Omega \nonumber 
\\
& = & 0
\end{eqnarray}
where, in the mesh system, the volume scalar is 
\begin{equation}
d \Omega \; = \; {\cal{L}} \, \sqrt{-g} \, 
\frac{\epsilon_{\alpha ' \beta ' \gamma ' \delta '}}{4 !} \, d \xi^{\alpha '} 
\, d \xi^{\beta '} \, d \xi^{\gamma '} \, d \xi^{\delta '} 
\end{equation}
the surface 1-form normal to the domain boundary is 
\begin{eqnarray}
\label {surface_one_form}
d(\partial \Omega)_{\beta} & = & {{\cal{L}}^{\alpha '}}_{\beta} \, 
d(\partial \Omega)_{\alpha '} \nonumber 
\\
& = & {\cal{L}} \, \sqrt{-g} \, 
{{\cal{L}}^{\alpha '}}_{\beta} \, \frac{\epsilon_{\alpha ' \beta ' \gamma ' \delta '}}{3 !} 
\, d \xi^{\beta '} \, d \xi^{\gamma '} \, d \xi^{\delta '} ~~~~~~
\end{eqnarray}
and $\epsilon$ is the flat-space Levi-Civita permutation tensor.  
{\em In the weak form, all terms involve only first-order derivatives of 
the variables $w^v$ with respect to the nodal coordinates.}  
Second order derivatives are generated by the ${\cal{W}}_{\inode , \beta}$ 
term which, after integration, differences the flux $T^{\beta}_{q}$ on each 
side of each node in a manner similar to a finite volume scheme.  

Note that, in order to generate the nodal equations in the 
interior of the domain, weights ${\cal{W}}_{\inode} ({\bm{x}})$ that vanish 
on the boundary always will be used.  Therefore, the first term in 
equation (\ref{weak_form}) --- the boundary term --- will always be zero.

\subsubsection{Second-order Equations:  The Galerkin Method}

The weighted residual method can be derived in a number of ways.  In early 
papers on finite element analysis, only linear problems were addressed and the 
nodal  equations were generated using a variational approach that maximizes 
the norm of the solution (\cite{zienk}).  This led to the form 
(\ref{strong_form}) with the shape function itself as the weight
\begin{equation}
{\cal{W}}^{G}_{\inode} ({\bm{x}}) \; = \; {\cal{N}}_{\inode} ({\bm{x}}) 
\end{equation}
This choice for ${\cal{W}}_{\inode}$ is called the Galerkin method and is 
especially useful for second-order equations.  For example, for the simple 
Poisson equation in one dimension ($w_{, x x} - \rho = 0$), with 
quadratic elements and uniform node spacing $\Delta x$, equation 
(\ref{weak_form}) generates the following nodal equation at the {\em central} 
node $\inode$ of each element
\begin{displaymath}
\frac{4 \Delta x}{3} [ \frac{w_{\inode-1} - 2 w_{\inode} + w_{\inode+1}}{\Delta x^2} \, 
- \, \frac{\rho_{\inode-1} + 8 \rho_{\inode} + \rho_{\inode+1}}{10} ] \; = \; 0
\end{displaymath}
which is similar to the finite difference form 
$(w_{\inode-1} - 2 w_{\inode} + w_{\inode+1}) / {\Delta x^2} \, - \, \rho_{\inode}
\, = \, 0$.  (Somewhat more complex 4th order difference equations are generated for nodes on 
element boundaries.)  In general the Galerkin weighted residual method generates derivatives 
similar to those expected in finite difference schemes (although in general geometry), 
but scalars and source terms are weighted averages of nodes surrounding $\inode$ 
rather than evaluated exclusively at $\inode$.  

An important property of the shape functions hints at a more fundamental 
interpretation of the weighted residual method that is not discussed usually 
in the engineering literature.  As the element volume $\delta \Omega$ 
approaches zero, the shape functions become good approximations to the Dirac 
delta function
\begin{equation}
\lim_{\delta \Omega \rightarrow 0} \, {\cal{N}}_{\inode} ({\bm{x}}) \; = \; 
k_{\inode} \, \delta \Omega \, \delta ({\bm{x}} - {\bm{x}}_{\inode})
\end{equation}
(no sum on $\inode$)
where $k_{\inode}$ is a scaling constant of order unity, but generally 
different for each node $\inode$.  Therefore, with the Galerkin method, the 
weights ${\cal{W}}_{\inode}$ in equation (\ref{strong_form}) are generalized 
approximations to $\delta ({\bm{x}} - {\bm{x}}_{\inode})$ which, when 
integrated over a {\em differential equation}, generate or ``pick out'' the 
corresponding {\em difference equation} near node $\inode$.  As the number of 
finite elements approaches infinity, the $\Im_{q \inode}$ more closely 
approximate the complete set of $\Re_{q}$ defined at all points in $\Omega$.  
Therefore, while originally derived for linear equations, the weighted 
residual method is valid for nonlinear problems as well.

\subsubsection{First-order Equations:  Petrov-Galerkin 
Schemes and ``Staggered Grids''}

The lack of a single, universally-applicable weighting function 
${\cal{W}}_{\inode} ({\bm{x}})$ is the main impediment against developing a 
truly general simulation code.  One must always know the order of differential 
equation being integrated.  For example, while the Galerkin scheme works well 
for second-order equations, it has the same pitfalls for first-order equations 
and fluid flow as centered-differencing schemes have in the FDM:  
leapfrogging, in which important terms in the equations do not depend on 
variables at the node at which the integral (\ref{weak_form}) is evaluated 
(($\partial w / \partial x)_{\inode} \approx (w_{\inode+1} - w_{\inode-1})/2 
\Delta x$), and two-point oscillations near shocks.  

Such problems can be addressed by using weighting functions other than the 
${\cal{N}}_{\inode}$. These are called Petrov-Galerkin schemes (\cite{hughes87}).  
For odd-order equations, functions that shift the peak of the weight away 
from node $\inode$ reduce or eliminate many of these problems.  This 
is the case for weights that are shape functions of twice the element 
interpolation order
\begin{equation}
\label {petrov_galerkin1}
{\cal{W}}^{PG1}_{\inode} ({\bm{x}}) \; = \; {\cal{N}}^{2 \wp}_{\inode + 
\frac{1}{2}} ({\bm{x}})
\end{equation}
and for those that are distorted by the shape function 
derivative
\begin{equation}
\label {petrov_galerkin2}
{\cal{W}}^{PG2}_{\inode} ({\bm{x}}) \; = \; {\cal{N}}_{\inode} \, \pm \, 
v^{\alpha} \, {\cal{N}}_{\inode , \alpha} ({\bm{x}})
\end{equation}
In the first case $\inode + \frac{1}{2}$ signifies a position in the mesh 
centered between nodes.  The functions ${\cal{W}}^{PG1}_{\inode}$ peak in 
between nodes and have almost the same effect as using a staggered grid does 
in the FDM.  
Both ${\cal{W}}^{PG1}_{\inode}$ and ${\cal{W}}^{PG2}_{\inode}$ generate 
non-leapfrogging differences (($\partial w / \partial x)_{\inode} \approx 
(w_{\inode+1} - w_{\inode})/ \Delta x$) and are still generally second-order 
accurate (or higher) as the scalar source terms are evaluated at the same 
place as the derivatives.

An upwinding scheme can be generated by setting $v^{\alpha} \propto u^{\alpha}$ 
and integrating only the advective terms with this weight.  However, this 
scheme has low order accuracy and is rather diffusive (\cite{hughes87}).  
Better methods for handling shocks are the van Leer scheme (\cite{vanleer79}), 
in which one enforces monotonicity in the gradient of the flux of a conserved 
quantity, and higher order Godonov schemes (\cite{cw84}; \cite{col90}) in 
which one applies the shock jump conditions within the element itself.  
Upwinding schemes in the FEM are a sub-field in themselves and are largely 
beyond the scope of this paper.

\subsection{Application of Boundary Conditions}

Not all of the nodal equations generated by (\ref{weak_form}) are useful.
For example, for each {\em second} order differential equation (one where a 
given ${\cal{T}}^{\beta}_{q}$ is a function of the gradient of at least one 
$w^v$) exactly ${\cal{K}}$ of the nodal equations (those integrated over shape 
functions peaking at boundary nodes on $\partial \Omega$) are meaningless or 
incomplete.  This is due to the absence of elements {\em beyond} the boundary 
needed to complete the integrals.  For {\em first}-order equations (ones where 
${\cal{T}}^{\beta}_{q}$ is, at most, a function of ${\bm{w}}$ and ${\bm{x}}$ 
only) those on only one portion of the boundary must be discarded ({\it e.g.}, 
on one side or surface).  
These ignored nodal equations must be replaced by exactly the same number of 
boundary or initial conditions.  These are generated in a weighted residual 
manner similar to that in equation (\ref{strong_form}) except that the 
integral and weighting functions now are evaluated on the boundary
\begin{equation}
\label {boundary_strong_form}
\mbox{\ss}_{r \knode} (w^v_{\jnode}) \; = \; - \int_{\partial \Omega} {\cal{W}}_{\knode} 
({\bm{x}}) \, f_{r} (w^v_{\jnode}) \, d({\partial \Omega})
\end{equation}
where $d({\partial \Omega})$ is the magnitude of the surface element
\begin{equation}
\label {abs_surface_one_form}
d({\partial \Omega}) \; = \; n^{\beta} \, d({\partial \Omega})_{\beta} 
\end{equation}
and the nodes $\knode$ lie on $\partial \Omega$.  
The number of equations, therefore, will remain equal to the number of 
unknowns, as is necessary for a well-posed problem.

\subsection{Numerical Aspects of the Method}

\subsubsection{Integration of the Weighted Residuals}

In the FEM the integrals in equations (\ref{weak_form}) and 
(\ref{boundary_strong_form}) often are performed numerically using Gaussian 
integration (\cite{as65}), with all sampled points ${\bm{x}}_g$ interior to 
element boundaries.  In practice, the integrals are calculated piecewise, 
element by element, with each element's contribution to the various integrals 
summed accordingly.  To accomplish this, one defines a {\em local} mesh 
coordinate system, referenced to the element center and parallel to the global 
$\xi^{\alpha '}$ system
\begin{eqnarray}
s^{\alpha ''}_{e} & \equiv & {\delta^{\alpha ''}}_{\alpha '} \, \frac{2 (\xi^{\alpha '} - 
\xi^{\alpha '}_{e})} { \Delta \xi^{\alpha '}_{e}}
\\
\frac{\partial \; \; \;}{\partial s^{\alpha ''}_{e}} & = & {\delta^{\alpha '}}_{\alpha ''}
\frac{\Delta \xi^{\alpha '}_{e}}{2} 
\frac{\partial \; \;}{\partial \xi^{\alpha '}} 
\\
-1 & \leq & s^{\alpha ''}_{e} \; \leq \; 1
\end{eqnarray}
where ${\bm{\xi}}_{e}$ is the position in mesh space of element $e$'s center, 
$\Delta \xi^{\alpha '}_{e}$ is the element width in direction $\xi^{\alpha '}$, 
and double primes refer to the local element system.  
The Lagrangian shape functions within each element take on a very simple form 
in this local system.  For linear interpolation in one dimension (with nodes 
lying at $s_{\tilde{1} e} = -1$ and $s_{\tilde{2} e} = +1$)
\begin{equation}
\label {linear_shape}
{\cal{N}} ''_{\ilocalnode e} (s_{e}) \; = \; 0.5 \, (1 \, + \, s_{\ilocalnode e} \, s_{e})
\end{equation}
For quadratic interpolation (with element nodes lying at $s_{\ilocalnode e} = -1, 0, +1$) 
\begin{equation}
\label {quadratic_shape}
{\cal{N}} '' _{\ilocalnode e} (s_{e}) \; = \; \left\{ \begin{array}{ll}
    0.5 \, s_{\ilocalnode e} \, s_{e} \, (1 \, + \, s_{\ilocalnode e} \, s_{e}) & 
    \mbox{($s_{\ilocalnode e} = \pm 1$)} 
\\
    (1 - s_{e}^2)                                             & \mbox{($s_{\ilocalnode e} = 0$)}
                                          \end{array}
                                   \right.
\end{equation}
and so on for higher order interpolation.  
(Equations [\ref{linear_shape}] and [\ref{quadratic_shape}] are the functions 
depicted in Figure \ref{shapes_fig}.)
Higher dimensional element shape functions are products of these in a manner 
similar to equations (\ref{lagrange_shapes}) for Lagrangian elements and 
(\ref{serendip_shapes}) for serendipitous elements.

Equation (\ref{weak_form}) then becomes, dropping the first term, as discussed 
earlier, and summing over elements and Gaussian integration points,
\begin{equation}
\label {gaussian_integ}
\Im_{q \inode} \; \approx \; - \, \sum_{e} \, \sum_{g} \omega_{g} 
\left[ {\cal{W}}_{\inode} ({\bm{x}}_{ge})_{ , \beta} T^{\beta}_{q} \, + \, 
{\cal{W}}_{\inode} ({\bm{x}}_{ge}) \, F_{q} \right] \delta \Omega_{e} 
\end{equation}
where 
\begin{equation}
\label {element_vol}
\delta \Omega_{e} \; = \; {\cal{L}} \, \sqrt{-g} \, \prod_{\alpha ' = 0}^{D-1} 
\Delta \xi^{\alpha '}_{e}
\end{equation}
is the element volume in the real-space coordinate system, 
$e = 1, ..., {\cal{E}}$ is the element number in the mesh, $g = 1, ..., G$ is 
the number of the Gaussian integration point within element $e$, and 
$\omega_{g}$ is the Gaussian weight at that point.  Similarly, equation 
(\ref{boundary_strong_form}) becomes
\begin{equation}
\label {boundary_integ}
\mbox{\ss}_{r \knode} \; \approx \; - \sum_{b} \, \sum_{g} 
\omega_{g} \, {\cal{W}}_{\knode} ({\bm{x}}_{gb}) \, f_{r} 
\, {n_{b}}^{\alpha '} \delta (\partial \Omega_{b})_{\alpha '} 
\end{equation}
where $b = 1, ..., {\cal{B}}$ is the boundary element number,  
\begin{equation}
\label {element_surface}
\delta (\partial \Omega_{b})_{\alpha '}  \; = \; {\cal{L}} \, \sqrt{-g} \, 
\prod_{\beta ' \neq \alpha '}^{D-1} \Delta \xi^{\beta '}_{b}
\end{equation}
is the surface 1-form on the $\alpha '$ boundaries, and 
$n_{b}^{\alpha '}$ is one of ${\delta^{\alpha '}}_{0 '}/\sqrt{-g_{0 ' 0 '}}$, 
${\delta^{\alpha '}}_{1 '}/\sqrt{-g_{1 ' 1 '}}$, {\it etc.}, depending 
on the boundary.

Because the residual weights $[{\cal{W}}_{\inode} ({\bm{x}}_{g}) 
= {{\cal{W}} ''}_{\inode} ({\bm{s}}_{g})]$, their derivatives, and the 
Gaussian weights $\omega_{g}$ are the same for all elements, they can be 
precomputed and stored prior to beginning the relaxation of the solution.  
The only quantities necessary to compute during the relaxation are the 
$T^{\beta}_{q}$ and $F_{q}$ at each ${\bm{x}}_{ge}$ interior integration point 
and the $f_{r}$ at each ${\bm{x}}_{gb}$ boundary integration point.  

Historically, the number of integration points $G$ used in each element is a function 
of the expected nonlinearity of the product ${\cal{W}}_{\inode} \Re_{q}$ with respect 
to position ${\bm{x}}$.  In engineering, this is usually of low order (linear or 
quadratic), but in astrophysics this product can vary with exponential order or higher.
Nevertheless, in practice, even with highly nonlinear functions, 
the author has had quite satisfactory results using the same number of 
integration points as nodes in each dimension ({i.e.}, $G = I$).  
Fewer than this (``underintegration'') reduces the order of accuracy or even 
can produce a singular matrix.  More than this (``overintegration'') does 
little to improve accuracy (of order unity improvements only) at great 
computational expense.

\subsubsection{Solution of the Simultaneous Nonlinear Equations:  The 
Multi-Dimensional Henyey Method}

For solving the $V {\cal{I}}$ nodal equations and boundary conditions, the author 
currently uses a standard multivariate Newton-Raphson technique, sometimes 
called the ``Henyey'' method in astrophysics (\cite{clayton68}; \cite{am65}).  
The $\Im_{q \inode}$ are linearized by expanding in a Taylor series about the 
solution $w^v_{\jnode}$, resulting in a matrix inversion problem
\begin{equation}
\label {newton_raphson1}
\left( \frac{\partial \Im_{q \inode}} {\partial w^v_{\jnode}} \right)^{[n]} \, 
\delta {w^v_{\jnode}}^{[n]} \; = \; - \Im_{q \inode}^{[n]}
\end{equation}
(no sum on the iteration number $[n]$) to solve for the corrections 
$\delta {w^v_{\jnode}}^{[n]}$ which need to be applied to the current guess 
to obtain a new [$n + 1$st] approximation to the solution
\begin{equation}
\label {newton_raphson2}
{w^v_{\jnode}}^{[n+1]} \; \approx \; {w^v_{\jnode}}^{[n]} \, + \, \delta 
{w^v_{\jnode}}^{[n]}
\end{equation}
In the engineering FEM ${\partial \Im_{q \inode}} / {\partial w^v_{\jnode}}$ 
is called the ``tangent stiffness'' matrix.  It has a length $V {\cal{I}}$ on 
a side and bandwidth $\sim \! V {\cal{I}}^{(1-1/D)}$.  At present GENRAL uses 
{\em direct} methods (Gaussian elimination with lower-upper decomposition) to 
solve equation (\ref{newton_raphson1}), repeatedly applying the corrections 
until the norm over all $v$ and $\jnode$ 
\begin{displaymath}
{\Delta}_{w} \equiv \| \delta {w^v_{\jnode}}^{[n]} / 
{w^v_{\jnode}}^{[n+1]} \|
\end{displaymath}
falls below a certain tolerance.

The tangent stiffness matrix need not be extremely accurate.  Indeed, when 
${\Delta}_{w} << 1$, the matrix need not be recomputed at all, with little 
impact on the rate of convergence and no impact on the accuracy of the 
solution.  Furthermore, the elements of the stiffness matrix in equation 
(\ref{newton_raphson1}) can be calculated using numerical differentiation, 
rather than writing out explicitly the partial derivatives of each equation 
with respect to each variable.  This eliminates the need to know the geometry 
of the mesh beforehand --- a feature important for multidimensional 
astrophysical structures.  Numerical differentiation of 
${\partial \Im_{q \inode}} / {\partial w^v_{\jnode}}$ is not necessarily more 
expensive than algebraic differentiation, especially if the the baseline 
residual integrals $\Im_{q \inode}^{[n]}$ are calculated only once for each 
matrix, and those partial derivatives known to be identically zero are not 
computed.

\subsubsection{Logarithmic Variables}

It is quite common for an astrophysical state variable --- {\it e.g.}, the 
density $\rho$ --- to vary by many orders of magnitude over $\Omega$.  
Therefore, in order to maintain the same relative accuracy over the domain, 
it may be necessary to solve for a much more slowly varying function, 
{\it e.g.}, $\tilde{\rho} \equiv \log_{10} (\rho)$.   In addition, for 
variables that can be positive or negative (like velocity) one may need a more 
complex function and its inverse
\begin{eqnarray}
\tilde{v} \; = \; {\rm slog}_{10} (v) & \equiv & S \, \log_{10} 
(1 \, + \, S \, v / v_{scale}) ~~~~~
\\
{\rm sdex} (\tilde{v}) & \equiv & S \, v_{scale} \, (10^{S \tilde{v}} - 1) 
\end{eqnarray}
where $S \equiv {\rm sgn} (v) = {\rm sgn} (\tilde{v})$ and 
$v_{scale}$ is a fixed scaling value for $v$.  
These ``scaled logarithmic variables'' are linear for $|v| << v_{scale}$ and 
logarithmic for $|v| >> v_{scale}$ and can be negative or positive.

Unfortunately, there is some loss of convenience and intuition in re-writing 
the equations in terms of these new, modified variables, especially if there 
are many of this nature.  Therefore, the following procedure has been devised 
in order that the differential equations still can be coded in their basic 
form ({\it i.e.}, using $\rho$ and $v$) while maintaining the accuracy of 
solving for $\log_{10} (\rho)$ and ${\rm slog}_{10} (v)$:
\begin{enumerate}
\item Each variable is flagged as being linear, logarithmic, or scaled logarithmic, 
and then stored as 
\begin{eqnarray*}
{\tilde{w}}^{v} & = & \ell(w^{v}) 
\\
& \equiv & \left\{ \begin{array}{ll}
    w^{v}                                              & \mbox{(if linear)} 
\\
    \log_{10} (w^{v})                                  & \mbox{(if logarithmic)} 
\\
    {\rm slog}_{10} (w^{v})                            & \mbox{(if scaled logarithmic)}
                                          \end{array}
                                   \right.
\end{eqnarray*}
\item Partial derivatives of the integrals in equation (\ref{newton_raphson1}) 
are calculated with respect to ${\tilde{w}}^{v}$.  However, when computing the 
functions $T^{\beta}_{q}$, $F_{q}$, and $f_{r}$, the stored variables and their 
gradients are re-exponentiated to their unmodified forms with 
\begin{eqnarray*}
w^{v} & = & \varepsilon({\tilde{w}}^{v}) 
\\
& \equiv & \left\{ \begin{array}{ll}
    {\tilde{w}}^{v}                                      & \mbox{(if linear)} 
\\
    10^{{\tilde{w}}^{v}}                                 & \mbox{(if logarithmic)} 
\\
    {\rm sdex} ({\tilde{w}}^{v})                         & \mbox{(if scaled logarithmic)}
                                          \end{array}
                                   \right.
\end{eqnarray*}
and
\begin{displaymath}
{w^v}_{, \beta} \; = \; \frac{{{\tilde{w}}^v}_{, \beta}} {\partial \ell(w^{v}) / \partial w^{v} }
\end{displaymath}
\end{enumerate}
With this scheme one can choose a variable to be logarithmic or not at runtime, or 
even switch its character during execution, without modifying the code.

\subsubsection{Pivoting}

Lower-upper decomposition techniques work well only when the matrix elements on 
the diagonal are decidedly non-zero.  That is, one must identify which equation 
$\Im_{q \inode}$ is ``for'' which variable $w^v_{\jnode}$.  The term 
{\em pivoting} refers to the exchange of rows and/or columns in the matrix to 
ensure that all elements on the diagonal are indeed large --- {\it i.e.}, that 
the trace of the stiffness matrix is a maximum.

{\em Local}, or partial, pivoting ensures that, at a given node, the 
correct physical equation is paired with the correct variable.  For example, 
in MHD computations, when the electrical conductivity is infinite, the current 
${\bm{\cal{J}}}$is determined by Maxwell's equations, not Ohm's law.  Or, in a 
hydrostatic star the momentum equation determines the pressure structure, 
while velocity is determined by energy or particle conservation.
In the standard partial pivoting algorithm one searches a matrix column for the 
largest element and switches the row of that element with the one presently on 
that column's diagonal (\cite{numrec}). 
When applied at a single node, this algorithm is very successful in 
automatically pairing equations and variables.  That is, the solution will be 
able to evolve from a dynamic state to a hydrostatic one without re-casting 
the equations or writing a new simulation code.

{\em Global} pivoting ensures that an equation $\Im_{q \inode}$ (integrated 
near node $\inode$) is applied at the correct node $\jnode$.  This is a 
more difficult task than local pivoting and is not easily automated in our 
case.  The correct identification is {\em not} always $\jnode = \inode$, 
especially for first-order equations, for which the answer is determined 
by where the boundary conditions are applied.  Fortunately, the global pivot 
of the matrix is a property that usually does not evolve with the simulation; 
it need be determined only once.  Therefore, an elaborate pre-pivoting scheme 
has been devised for GENRAL, in which, based only on the equation order and 
location of boundary conditions, shape function coefficients are identified 
with nodal equations, useless nodal equations are discarded, and boundary 
conditions are inserted.  Local pivoting can still shift emphasis for a given 
variable from one differential equation to another, but the overall global 
identification of integrals (\ref{weak_form}) with nodes remains fixed.

\section{The Finite Element Method with Nodal Coordinates as Part of the Solution}

In most multi-dimensional astrophysics problems, the grid chosen initially will 
not be a good match for the final structure, due to a poor fit to the object 
boundary or poor resolution in areas of rapid gradients.  
Therefore, the coordinates and/or quantity of nodes should be changed as the 
solution converges in order to get a better fit (``adaption'').  In addition, 
one needs to allow for the mesh at one time step to be different from that at 
the previous time step (grid motion).  

\subsection{Adaptive Gridding}

Adaptive gridding, as used here, means modifying the mesh spacing 
in order to achieve greater accuracy or stability 
{\em without changing the total number of nodes or the topology of the mesh}.  
One-dimensional stellar structure models use a form of adaptive gridding 
as they utilize a mass coordinate rather than radius, allowing the radius 
of each mass zone, including the outer stellar radius, to expand or contract 
depending on the current state of the star.  Our general multidimensional 
adaptive gridding scheme takes the same form as equation (\ref{general_eq}) 
except that it is written in mesh space
\begin{equation}
\label {adapt1}
{{\cal{A}}^{\alpha ' \beta '}}_{, \beta '} \; = \; 0
\end{equation}
where ${\cal{A}}^{\alpha ' \beta '}$ is the adaptive gridding tensor.  
Currently the author is using a diagonal form 
\begin{equation}
\label {adapt2}
{\cal{A}}_{\alpha ' \alpha '} \; = \; [ f_v \, (C_1 \, {\bm{e}}_{\hat{\alpha} '} 
\cdot {\bm{\nabla}} w^v \, - \, \frac{< \! \! \Delta w^v \! \! >}{\Delta x^{\alpha '}} )
\, + \frac{C_2}{\Delta x^{\alpha '}} ] 
\end{equation}
which ties the mesh spacing to the local gradient of the state variables. 
Note the sum over $V$ variables; $f_v$ is a vector of 1's and 0's that, at run 
time, selects those state variables to which the grid should be adapted, 
${\bm{e}}_{\hat{\alpha} '}$ is a unit vector in spacetime along the local mesh 
direction $\xi^{\alpha '}$ 
\begin{displaymath}
[{\bm{e}}_{\hat{\alpha} '}]^{\alpha} = \frac{{{\cal{L}}^{\alpha}}_{\alpha '}}
{ |g_{\mu \nu}{{\cal{L}}^{\mu}}_{\alpha '} \, {{\cal{L}}^{\nu}}_{\alpha '}|^{1/2} }
\end{displaymath}
(no sum on $\alpha '$), $< \! \! \Delta w^v \! \! >$ is the average change in 
$w^v$ along the mesh direction $\xi^{\alpha '}$ (determined from the current 
solution for $w^v$), $\Delta x^{\alpha '}$ is a measure of the local linear 
mesh spacing along an element edge 
\begin{eqnarray}
\Delta x^{\alpha '} & = & \frac{\Delta \xi^{\alpha '}}
{{\bm{e}}_{\hat{\alpha} '} \cdot {{\cal{L}}^{\alpha '}}_{\alpha}} \nonumber
\\
& = & |g_{\alpha ' \alpha '}|^{1/2} \, \Delta \xi^{\alpha '}
\end{eqnarray}
(again, no sum on $\alpha '$), and $C_1 \approx 0.2$ is a constant that 
regulates the strength of the gradient term.  
Equations (\ref{adapt1}) provide four constraints on the nodal coordinate 
values, allowing the mesh spacing along $\xi^{\alpha '}$ to decrease in 
regions of high gradients in the variables, but to be uniform otherwise.  
Note that the adaptive gridding equations are not the coordinate conditions 
needed to complete the description of the metric (see Appendix A).
They simply move nodal positions around in an 
already-determined metric.  $C_2$ is a very small constant 
($\approx 10^{-10}$), such that in the absence of adaptive gridding 
({\it i.e.}, all $f^v = 0$) and with the proper boundary conditions, the 
mesh will assume an appropriate curvilinear character with uniform spacing 
$\Delta x^{\alpha '}$ in each $\xi^{\alpha '}$.

\subsection{Moving Grids and the Advective Derivative}

Adapting the grid to local conditions changes the spatial and time coordinates 
of the nodes.  For an observer traveling along $\xi^{0'}$ this 
produces what appears to be motion of the grid through space.  Now, 
computational fluid dynamics on stationary, uniformly-spaced meshes is a fairly 
complex field in itself.  On arbitrarily-spaced and moving grids the resulting 
mixed Lagrangian-Eulerian hydrodynamic equations might seem intractable.  
However, with the FEM the exact opposite is true.  {\em If the elements have an 
extent in time as well as in space, then the inverse isoparametric transformation 
will automatically take into account the complicated effects of differencing 
the fluid equations with respect to a moving grid.}  It is unnecessary, and 
indeed incorrect, to attempt to include the grid velocity in the differential 
equations.

As an example, consider the non-relativistic total advective derivative in flat spacetime 
\begin{eqnarray}
\frac{D~}{Dt} & = & \frac{\partial ~}{\partial t} \, + \, v^i 
\frac{\partial ~}{\partial x^i} \nonumber
\\
\label {advect_der}
& = & \frac{\partial \xi^{\alpha '}}{\partial t \;} 
\frac{\partial ~}{\partial \xi^{\alpha '}}
\, + \, v^i \frac{\partial \xi^{\alpha '}}{\partial x^i} 
\frac{\partial ~}{\partial \xi^{\alpha '}}
\end{eqnarray}
where $v^i$ is the three-velocity of fluid flow.  Now, let $v^i_g$ be a 
grid three-velocity such that ${\bm{e}}_{\xi^{i'}}$ remains parallel to ${\bm{e}}_{x^i}$,
although ${\bm{e}}_{\xi^{0'}}$ can make an angle ($\tan \theta = |v_g| \, \Delta t/ 
|\Delta x|$) with ${\bm{e}}_t$.  (These are the conditions \cite{lw} placed on 
their moving grid in their Lagrangian-Eulerian MHD calculations.)  The isoparametric 
transform and its inverse then are 
\begin{eqnarray}
\label {isoparm}
\frac{\partial x^{\beta}}{\partial \xi^{\alpha '}} 
   & = &  \left\{ \begin{array}{ll}
   \frac{\Delta x^{\beta}} {\Delta \xi^{\alpha '}}    &  \mbox{(if $\beta = \alpha '$)}
\\
   \frac{v^i_g \, \Delta t}{\Delta \xi^{0'}}           &  \mbox{(if $\beta = i$; $\alpha ' = 0$)} 
\\
   0                                                  &  \mbox{(otherwise)}
                                          \end{array}
                                   \right.
\\
\label {isoparm_inv}
\frac{\partial \xi^{\alpha '}}{\partial x^{\beta}} 
   & = & \left\{ \begin{array}{ll}
  \frac{\Delta \xi^{\alpha '}} {\Delta x^{\beta}}     &  \mbox{(if $\alpha ' = \beta$)}
\\
   - \frac{v^i_g \, \Delta \xi^{i'}}{\Delta x^i}       &  \mbox{(if $\alpha ' = 0$; $\beta = i = i'$)}
\\
   0                                                  &  \mbox{(otherwise)}
                                          \end{array}
                                   \right.
\end{eqnarray}
(no sum on $\beta$ or $i$).  Substituting equation (\ref{isoparm_inv}) into 
(\ref{advect_der}) one obtains
\begin{equation}
\label {moving_advect_der}
\frac{D~}{Dt} \; = \; \frac{\Delta \xi^{0'}}{\Delta t} \frac{\partial ~}{\partial \xi^{0'}} 
\, + \, (v^i - v^i_g) \,  \frac{\Delta \xi^{i'}}{\Delta x^i} 
\frac{\partial ~}{\partial \xi^{i'}}
\end{equation}
The first term in equation (\ref{moving_advect_der}) is the apparent time 
derivative (along $\xi^{0'}$) at a given node in the grid frame 
while the second is simply the moving advective derivative 
$({\bm{v}}-{\bm{v}}_g) \bm{\cdot \nabla}$.  
Thus the isoparametric transform reproduces the Lagrangian-Eulerian equations 
of LeBlanc \& Wilson under the same conditions.

\subsection{Adaptive Mesh Refinement}

The purpose of adaptive mesh refinement (AMR) is similar to that of adaptive 
gridding:  to increase resolution in a local region of the mesh.  However, in 
the case of AMR the numbers of elements and nodes usually increase as the 
calculation proceeds.  
When the element mesh extends into the time dimension, AMR makes it possible 
to treat efficiently the propagation of particularly important wave phenomena.  
Rather than subdivide the entire domain finely, one does so only for the region 
of spacetime near the null geodesics along which the phenomenon propagates. 
The Courant condition (Appendix B) prescribes one 
form of AMR, as it places an upper limit on the temporal mesh spacing and, 
therefore, a lower limit on the number of elements necessary to solve a 
problem.  If the Courant condition is not satisfied in a local region, the 
number of elements in $\xi^{0'}$ will have to be increased.  

In the FEM AMR can be achieved by subdividing some of the elements, usually 
by a power of two in a given dimension, and then rebuilding the 
interpolation grid of nodes within the newly-gridded subdomain.  There 
are two matters of concern in this case:  1) the subdivision must be done by 
elements, not nodes, regardless of the final order of the interpolation grid 
used within each element, and 2) larger elements bounding the more finely 
subdivided region must have their internal interpolation scheme, and hence 
their nodal shape functions, modified by the addition of new boundary nodes.
That is, these modified elements must be serendipitous elements.  This 
ensures that the functions $w^v ({\bm{\xi}})$ and $x^{\alpha} ({\bm{\xi}})$
are continuous across element boundaries and that there are no spatial 
gaps or ``hanging nodes''.  This second requirement can significantly increase 
the complexity of the mesh and the number of element types for which 
quantities need to be pre-computed and stored.  

Recently, however, the astrophysical community has been embracing an 
alternative, multi-{\em level} approach to AMR (\cite{tru98}; \cite{norman98}). 
A region of space is refined not by 
subdividing the grid cells themselves, but by applying separate, and 
successively more refined, grids at the same location and with some nodes in 
common between each level.  This hierarchical approach eliminates the hanging 
node problem without resorting to defining many new serendipitous element 
types: each grid level is subdivided with standard linear or quadratic 
elements.  This technique also fits naturally into a multigrid iterative 
scheme for solving the coupled nodal equations.

\section{Tests and Examples}

Below are presented some tests of the code GENRAL on problems with known 
solutions.  At the 
present time the code uses the multi-dimensional Henyey technique to solve the 
difference equations it generates.  As discussed earlier, this approach has 
severe computer time and memory limitations.  Therefore, all the examples 
involve a much smaller number of elements than the millions that one would use 
in a typical astrophysical simulation.  
(Generally, the full code running on a desktop workstation is limited to 4096 
elements or 6561 nodes total [$64^2$, $16^3$, or $8^4$ linear elements or 
$32^2$, $8^3$, or $4^4$ quadratic elements]).
Nevertheless, the tests serve to demonstrate the unique features 
of the FEM, including its ability to solve nonlinear astrophysics-like problems 
in multidimensional, arbitrary curvilinear coordinate systems and to achieve 
high accuracy in the solution by employing higher order interpolation, adaptive 
gridding, and logarithmic variables.

\subsection{Simple Tests}

\subsubsection{Tests of the Interpolation Scheme}

The first tests of the method involve no differential equations.  Instead, the 
integrands in equations (\ref{gaussian_integ}) and (\ref{boundary_integ}) are 
set to unity so that only domain volume and surface area are computed.  The 
results are then compared with known solutions for various shapes (rectangular 
hyper-solids, hyper-spheres, and ellipsoids), different element interpolation 
schemes (linear [$\wp=1$] and quadratic [$\wp=2$]), different numbers of elements 
$\aleph$ in each dimension (1,2,4,8,16), and different numbers of integration 
points $G$ ($\wp^D$ and $(\wp+1)^D$).  This procedure is more than 
simply a verification that the numerical integrations work.  It also exercises 
the coordinate transformations and the ability to form proper volume and 
surface elements in fairly arbitrary curvilinear situations.  

Table 1 summarizes the results, normalizing to grids with nine nodes in each 
dimension.\footnote{Given that the method is element-based, one could compare equal 
numbers of linear and quadratic {\em elements} in each dimension.  In that case, the errors 
in the right-hand column of Table 1 would be a factor of 16 smaller.  However, 
performing the comparison in the latter manner would make it difficult to tell which part 
of the improvement for quadratic elements is due to the interpolation scheme alone 
and which is due to the increase in the number of nodes.}  In this case (but, 
unfortunately, {\em not} for the differential equation tests below) both choices for $G$ 
produce the same results:  for linear elements the error ${\sf{E}}_{\Omega} \equiv 
(\Omega_{\aleph} - \Omega)/\Omega$ varies as $\Delta x^2 \propto (\aleph \wp - 1)^{-2}$, 
while quadratic elements produce fourth-order errors (${\sf{E}}_{\Omega} \propto \Delta x^4 
\propto (\aleph \wp - 1)^{-4}$).  Results for surface area integration are similar, although 
the actual fractional error is about a factor of two smaller.  Results for other 
ellipsoidal objects are also similar.

\begin{table}
\begin{tabular}{ccc} 
                    &                       &                       \\
                    &                       &                       \\
\multicolumn{3}{c}{TABLE 1}                                                                 \\
\multicolumn{3}{c}{Error Results for Volume Integration of $D$-Spheres\tablenotemark{a} }\\
\multicolumn{3}{c}{(Meshes with 9 {\em Nodes} in Each Dimension)}   \\
\hline
\multicolumn{3}{c}{~~~~~~~~~~ Fractional Error for Element Type}    \\
~~~~~ $D$ ~~~~~     & Linear\tablenotemark{b} & Quadratic\tablenotemark{b} \\
\hline \hline
                    &                       &                       \\
 ~~~~~ 2 ~~~~~      & $-6.4 \times 10^{-3}$ & $-4.9 \times 10^{-5}$ \\
 ~~~~~ 3 ~~~~~      & $-2.0 \times 10^{-2}$ & $-9.0 \times 10^{-5}$ \\
 ~~~~~ 4 ~~~~~      & $-3.7 \times 10^{-2}$ & $-2.2 \times 10^{-4}$ \\
\end{tabular}
\tablenotetext{a}{Surface area results have a similar scaling, but with a factor of 
two better accuracy}
\tablenotetext{b}{Accuracy for linear elements varies with $\Delta x^2$ and quadratic elements 
as $\Delta x^4$}
\end{table}

\begin{table}
\begin{tabular}{cccc} 
                    &                       &                      &                      \\
                    &                       &                      &                      \\
\multicolumn{4}{c}{TABLE 2}                                                               \\
\multicolumn{4}{c}{Error Results for Cartesian Grid Tests}                                \\
\multicolumn{4}{c}{(Meshes with 9 {\em Nodes} in Each Dimension)}                         \\
\hline 
\multicolumn{4}{c}{~~~~~~~~~~ Normalized L2 Error for Element Type} \\
  ~~~~~ $n$         &  $D$                  & Linear\tablenotemark{a} & Quadratic\tablenotemark{a} \\
\hline \hline
                    &                       &                      &                      \\
  ~~~~~ 1           &  1                    & \tablenotemark{b}    & \tablenotemark{b}    \\
  ~~~~~ 1           &  2                    & $2.5 \times 10^{-3}$ & $1.2 \times 10^{-3}$ \\
  ~~~~~ 1           &  3                    & $1.4 \times 10^{-3}$ & $4.0 \times 10^{-4}$ \\
  ~~~~~ 1           &  4                    & $9.9 \times 10^{-4}$ & $1.8 \times 10^{-4}$ \\
                    &                       &                      &                      \\
  ~~~~~ 2           &  1                    & $5.8 \times 10^{-3}$ & \tablenotemark{b}    \\
  ~~~~~ 2           &  2                    & $3.3 \times 10^{-3}$ & \tablenotemark{b}    \\
  ~~~~~ 2           &  3                    & $2.3 \times 10^{-3}$ & \tablenotemark{b}    \\
  ~~~~~ 2           &  4                    & $1.8 \times 10^{-3}$ & \tablenotemark{b}    \\
                    &                       &                      &                      \\
  ~~~~~ 3           &  1                    & $1.2 \times 10^{-2}$ & $1.2 \times 10^{-3}$ \\
  ~~~~~ 3           &  2                    & $6.1 \times 10^{-3}$ & $6.2 \times 10^{-4}$ \\
  ~~~~~ 3           &  3                    & $4.0 \times 10^{-3}$ & $4.0 \times 10^{-4}$ \\
  ~~~~~ 3           &  4                    & $3.0 \times 10^{-3}$ & $2.9 \times 10^{-4}$ \\
                    &                       &                      &                      \\
  ~~~~~ 4           &  1                    & $2.1 \times 10^{-2}$ & $3.1 \times 10^{-3}$ \\
  ~~~~~ 4           &  2                    & $1.0 \times 10^{-2}$ & $1.6 \times 10^{-3}$ \\
  ~~~~~ 4           &  3                    & $6.4 \times 10^{-3}$ & $1.0 \times 10^{-3}$ \\
  ~~~~~ 4           &  4                    & $4.6 \times 10^{-3}$ & $7.2 \times 10^{-4}$ \\
\end{tabular}
\tablenotetext{a}{Accuracy for linear elements varies with $\Delta x^2$ and quadratic elements 
as $\Delta x^3$}
\tablenotetext{b}{Exact solution obtained for one-dimensional first-order problems and for 
all second-order problems with quadratic elements}
\end{table}

\subsubsection{Fixed Cartesian Grid Tests:  Poisson's Equation in One to Four Dimensions}

The second set of tests involves solving a differential equation in up to four 
dimensions, but on a regular Cartesian (not Minkowskian) grid of dimension $D$.  The equation 
used is Poisson's equation 
\begin{equation}
\label {poisson_eq}
w_{, \alpha \alpha} \; - \; \rho \; = \; 0
\end{equation}
with a known solution
\begin{equation}
w_0 \; \equiv \; r^n
\end{equation}
where $r \equiv \sqrt{x^2 + y^2 + z^2 + t^2}$ 
is the radial distance from the origin and $0 < x^{\alpha} < 1$.  The components of the 
generalized equation (\ref{general_eq}) are then
\begin{eqnarray}
\label {simple_stress}
T^{\beta} & = & w_{, \beta} 
\\
\label {simple_force}
F & = & \nabla^2 w_0 \; = \; n \, (n+D-2) \, r^{(n-2)}
\end{eqnarray}
with Dirichlet conditions on the entire boundary
\begin{equation}
f = w - [r(\partial \Omega)]^n
\end{equation}
where $r(\partial \Omega)$ is the expression for $r$ evaluated on the boundary.  
This test exercises the code's ability to solve equations 
(\ref{gaussian_integ}) and (\ref{boundary_integ}), but does not test the 
coordinate transformations.

Results of the fixed grid tests are given in Table 2, which shows how the 
accuracy of the solution in different dimensions, determined by the normalized 
``L2'' error norm
\begin{displaymath}
{\sf{E}}^{L2}_w \; \equiv \; \frac{\left [ \int_{\Omega} (w - w_0)^2 \, d \Omega \right ]^{1/2}}
{\left [ \int_{\Omega} w_0^2 \, d \Omega \right ]^{1/2}}
\end{displaymath}
varies with number of nodes and elements used.  Two differences from the volume 
and surface area integration tests are worth noting.  Firstly, the solution errors 
for quadratic elements are third-order accurate (${\sf{E}}^{L2}_w \propto \Delta x^3 \propto 
(\aleph \wp - 1)^{-3}$), not fourth-order.  Secondly, underintegration ($G = \wp^D$) 
does not work.  {\em $G$ always must be equal to or greater than $(\wp+1)^D$} ({\it i.e.}, 
$\ge 2^D$ for linear elements and $\ge 3^D$ for quadratic elements) in order to generate 
the proper second-order differences in the Laplacian.  Underintegration, at best, reduces 
the accuracy of the solution by one order.  At worst, it destroys nearest neighbor 
differencing, producing leap-frogged differences, and can lead to no solution at all.
``Iso-integration'' ($G=I$) is probably sufficient for any equation of the form
(\ref{general_eq}), but it should be checked in each circumstance to be certain.

\subsection{Adaptive and Curvilinear Grid Tests}

The third set of tests exercises nearly all the features of the code in order 
to obtain a solution to a rather pathological Poisson problem --- a Fermi-Dirac-like 
function
\begin{equation}
w_0 \; \equiv \; \frac{1}{ e^{f(r-0.5)} + 1 }
\end{equation}
with a cold temperature of 1/f = 0.02.  Such sudden exponential drops in the solution 
are common at stellar core-halo boundaries or stellar surfaces and are difficult to 
resolve accurately without a large number of nodes or a variable change (to optical 
depth, for example).  For this demonstration the conservative form for 
the stress and force terms (in up to four Cartesian dimensions) has been chosen
\begin{eqnarray}
\label {adaptive_stress}
T^{\beta} & = & w_{, \beta} + f w \frac{a_{\beta}^2 \, x^{\beta}}{r [1 + e^{-f(r-0.5)}]}
\\
\label {adaptive_force}
F & = & 0
\end{eqnarray}
(no sum on $\beta$) where $a_{\beta} = (1, a, b, c)$ are constants and 
\begin{eqnarray*}
r & \equiv & \sqrt{x^2 + a^2y^2 + b^2z^2 + c^2t^2}
\end{eqnarray*}
(For example, in one dimension, $a=b=c=0$; in two dimensions, $b=c=0$; and so on.)
And, as one is still interested at this stage in testing the FEM machinery and not 
the astrophysical viability of the code, once again simple Dirichlet 
boundary conditions are employed, so equation (\ref{simple_bc}) becomes 
\begin{equation}
f = w - \frac{1}{ \exp[ f (r(\partial \Omega) - 0.5)] \, + \, 1 }
\end{equation}
rather than, for example, a multipole expansion of the interior solution.  The above 
conservative form (\ref{adaptive_stress} - \ref{adaptive_force}) was chosen in favor 
of other forms (such as \ref{simple_stress} - \ref{simple_force}) 
because its solutions are particularly accurate for a small number of nodes and suitable 
for demonstrating adaptive gridding and the use of logarithmic variables.

Figure \ref{1dpoisson_fig} shows the solution of this Poisson problem in one dimension as one 
applies successively more features of the code.  The top panels of Figures \ref{1dpoisson_fig} 
show standard fixed, equally-space grids of 8 and 16 linear elements respectively.  Some 
improvement in accuracy can be obtained by doubling the resolution, 
but this incurs additional storage and computational expense.  Turning on 
the adaptive gridding equations (\ref{adapt1}), however (middle panels), 
significantly improves accuracy for the same number of elements.  
This also demonstrates one aspect of the ispoarametric transformation:   variable node 
spacing.  A closer examination of this more accurate solution, however (Figure \ref{1dpoisson_fig}, 
bottom left), shows very large relative errors in the log when $w << 1$.  Nevertheless, 
these can be overcome easily, without re-writing and re-coding the equations, 
as the bottom right panel of Figure \ref{1dpoisson_fig} shows.  When $w$ is identified as a 
logarithmic variable, rather 
than linear, the solution remains accurate over eleven orders of magnitude. 
(Note the different node spacing in the bottom panels, with the grid adapting 
to $w$ on the left and $\tilde{w}$ [$= \log_{10} w$] on the right.)

Figure \ref{2dpoisson_fig} shows a similar development for a two-dimensional Poisson problem 
where the Fermi-Dirac surface has an elliptical ratio of 2:1 ($a = 2$).  
The first three panels demonstrate the errors possible in locating the surface if 
the proper coordinate geometry is not used.  Additional improvement in accuracy can 
be obtained by using adaptive gridding (middle right panel).  However, this solution 
for $r > 0.5$ suffers 
the same oscillatory errors seen in the one-dimensional case (bottom left), which again 
can be eliminated by identifying $w$ as logarithmic (bottom right).

It is important to note that, in all of the solutions displayed in Figures \ref{1dpoisson_fig} 
and \ref{2dpoisson_fig}, 
{\em no explicit curvilinear coordinate system is used}.  The coordinates of the 
grid points (whether fixed or part of an adaptive gridding solution) are stored only 
as $x_{\inode}$ and $y_{\inode}$, not as $r_{\inode}$ and $\theta_{\inode}$, for example, 
and yet are still fully arbitrary (subject to the Jacobi condition).  
The Poisson equation is written only in terms of $x$ and $y$ as well.  Of course, 
the derivatives are still calculated using the coordinate grids shown, but then they 
are immediately transformed to $(x,y)$-space using the transformations (\ref{isoparm_general}) 
and (\ref{isoparm_general_inv}).  Thus, with these new techniques the grid can 
be moved around to obtain a more accurate solution while the physical equations remain 
coded in the same very simple form.

The Fermi-Dirac Poisson tests were used to determine a good value for the adaptive 
gridding constant in equation (\ref{adapt2}).  Several dozen models like those in 
Figures \ref{1dpoisson_fig} 
and 3 were computed for different values this parameter.  It was found that the 
accuracy improved by factors of $3-10$ as $C_1$ was increased from $0$ to 
$0.2$, but beyond this point the accuracy did not improve much.  In fact, for values 
much greater than $0.2$, the models became unstable, often not converging.  Therefore, 
$C_1 = 0.2$ was chosen as a semi-universal value in the adaptive gridding equation.  
It has proven to be useful both in the Fermi-Dirac tests in Figures \ref{1dpoisson_fig} 
and \ref{2dpoisson_fig} and in the 
stellar structure models below.

One important point about adaptive gridding should be mentioned.  As currently 
implemented, the technique is rather volatile and unstable.  Unless great care 
is taken, iterations with adaptive grids often diverge, violating the Jacobi condition 
in the process.  In the case where a {\em single} solution to a steady state problem is sought, 
sometimes less CPU time cost will be incurred by subdividing the mesh more finely or 
using quadratic elements, rather than using adaptive gridding techniques.  On the other 
hand, when many thousands of successive models are to be computed, as is the case 
for evolutionary problems, each newly-converged model will be a good initial approximation 
to the next evolutionary state, yielding convergence for each time step in only a few 
iterations.  In this case, the amount of time spent converging the first adaptively-gridded 
model will be a small cost compared to the CPU time adaptive gridding saves over the course of 
the evolution by using a smaller number of elements and nodes to obtain the same high level 
of accuracy.

\subsection{Stellar Structure Tests:  Polytropic Stars}

The fourth series of tests adds the ability to solve a {\em coupled} set of both 
first- and second-order partial differential equations.  It also demonstrates the 
use of $\sqrt{-g}$ to solve a problem in which the {\em basic} coordinate 
system (not just the grid) is curvilinear, due to the assumption of symmetry 
conditions involving a coordinate direction orthogonal to the computational domain.  

\subsubsection{Spherical Polytropes in One Dimension}

In one dimension the equations for polytropic stellar structure are hydrostatic 
equilibrium (Euler's equation with zero velocity), Poisson's equation for gravity, 
and the polytropic equation of state
\begin{eqnarray}
\frac{d p}{d r} \, + \, \rho \, (n+1) \, \frac{d w}{d r} & = & 0
\\
\frac{1}{r^2} \frac{d \;}{d r} \left ( r^2 \, \frac{d w}{d r} \right ) 
\; - \; \rho & = & 0
\\
p & = & \rho^{1 + 1/n}
\end{eqnarray}
Pressure $p$ and density $\rho$ are unity at the stellar center and zero at its 
surface, and $r$ is the spherical radius coordinate.  The polytropic index $n$ is a 
measure of the hardness of the equation of state; the factor $(n+1)$ in the 
hydrostatic equilibrium (HSE) equation is a normalization constant for the gravitational 
potential $w$.  

In semi-analytic treatments, the hydrostatic equilibrium equation is multiplied 
by $r^2/\rho$, differentiated with respect to the radius $r$, and combined with 
Poisson's equation to give a single second-order equation.  
Following this procedure here, however, tests only our ability to solve that 
simple equation and not much else.  A slightly stronger test of the 
method would be to leave the system as a set of coupled equations and identify 
\begin{eqnarray}
T^r_{p} & = & p_{, r} \, + \, \rho \, (n+1) \, w_{, r} \nonumber
\\
F_{p} & = & 0 \nonumber
\\
\label{oned_poly_w_t}
T^r_w & = & w_{, r}
\\
\label{oned_poly_w_f}
F_w & = & \rho
\\
\label{oned_poly_rootg}
\sqrt{-g} & = & r^2
\end{eqnarray}
with boundary conditions at $r=0$ of $p = 1$ and $p_{, r} = 0$.  Note the need for 
a non-unit volume element of 
$r^2$ in order to form the proper divergence.  A basic curvilinear coordinate system 
must be used because of the spherical symmetry assumed in 
directions orthogonal to ${\bm{e}}_r$.  

Unfortunately, the above approach is still unsatisfactory, because it cleverly casts 
HSE as a second-order equation, avoiding the first-order equation problems discussed 
earlier.  Also, as shown below, it does not lend itself to generalization to two or 
more dimensions.  To address these issues, the following alternate identification of 
the stress and force terms for the pressure equation has been chosen 
\begin{eqnarray}
\label{oned_poly_p_t}
T^r_{p} & = & 0 
\\
\label{oned_poly_p_f}
F_{p} & = & p_{, r} \, + \, \rho \, (n+1) \, w_{, r}
\end{eqnarray}
which casts HSE as a first-order equation (with only the boundary condition $p = 1$ at 
$r = 0$).  Boundary conditions on the potential $w$ are $w_{, r}=0$ at the stellar center 
and $w= -r_{surface}/r$ at the stellar surface.  In addition, the condition $p = p_{s} 
\equiv 10^{-4}$ is applied at the surface on the adaptive gridding equation to determine the 
stellar radius.

Figure \ref{1dsph_poly_fig} shows an $n=1$ polytrope in one dimension, which has the analytic solution
\begin{displaymath}
\rho \; = \; \frac{\sin r}{r} ~~~~~~~~  w \; = \; -\rho - 1 ~~~~~~~~ p \; = \; \rho^2
\end{displaymath}
The left panel uses only the Galerkin method, while the right two panels show 
the solution using the Petrov-Galerkin scheme in equation (\ref{petrov_galerkin2}) to 
integrate the HSE equation.  The need to treat first-order equations differently from 
second-order ones is clearly evident.  The leapfrogging first-order differences produced 
by the Galerkin method not only display point-to-point oscillations, they also miss an implicit 
boundary condition ($p_{, r}=0$ at $r=0$, implied by $w_{, r}=0$ and equation \ref{oned_poly_p_f}).
The Petrov-Galerkin scheme improves the accuracy considerably, and with adaptive gridding 
remains roughly second order accurate for linear elements and third order for quadratic.

At the present time no robust, automated method 
for determining the order of the differential equation has been developed.  The 
code must be told explicitly not only the order of each equation but also the location(s) 
of the boundary conditions.  While this is the greatest obstacle to producing a truly 
general continuum simulation code to solve all types of equations, it is a relatively 
modest amount of effort compared with writing a new code for each problem.

\subsubsection{Rotating Polytropes in Two Dimensions}

Treatment of a uniformly-rotating polytropic star in two dimensions poses several problems.  
Firstly, it appears to be an overdetermined system
\begin{eqnarray}
\label {density_eq}
\nabla p \, + \, \rho \, (n+1) \, \nabla w \, - \, \rho R E_{rot} \, {\bm{e}}_R & = & 0
\\
\frac{1}{R} \frac{\partial}{\partial R} \left ( R \frac{\partial w}{\partial R} \right )
\, + \, \frac{\partial^2 w}{\partial Z^2} \; - \; \rho & = & 0 ~~~~~
\label {poissons_eq}
\\
p & = & \rho^{[1 + 1/n]} \; \; ~~~~~
\end{eqnarray}
(where $E_{rot} \equiv {\omega}^2/4 \pi G \rho$ is the normalized rotational energy per unit mass, 
$\omega$ is the uniform angular velocity of stellar rotation, and $R$ and $Z$ are the 
cylindrical radius and axis coordinates) 
with four equations but only three unknowns.  In normal astrophysical situations 
this dilemma will not arise, as the fluid equations plus the conservation of 
energy are a well-posed problem.  However, it occurs here because two unknowns 
($v_{R}$ and $v_{Z}$) and only one equation (conservation of mass) have been removed 
from the full set.  The trick is to convert the two redundant equations for HSE 
(\ref{density_eq}) into only one.  

One possible solution is to form a single second-order equation by taking the 
divergence of the HSE equation, similar to the standard semi-analytic approach.  
However, while it works fine in one dimension, this approach is 
unstable to two-dimensional perturbations when both the Dirichlet and Neumann boundary 
conditions are applied on the same ($r=0$) surface, leaving the stellar 
surface free.  
Setting $p = p_{s}$ at the surface does not help either; in this case the mesh 
must become adaptive, and this constraint must be used as a boundary condition 
on the adaptive gridding equations, not on HSE.  Moving the Neumann condition 
to the stellar surface is a better approach, but difficult to apply for more 
complex problems ({\it e.g.}, rotating polytropes).

An approach that does work is to {\em project} the HSE equation along a direction 
in which $p$ and $w$ have significant gradients.  The projection direction 
need not be along ${\bm{e}}_{p} \equiv -\nabla p / |\nabla p|$; ${\bm{e}}_{r}$ 
appears sufficient, even when the polytrope is rotating rapidly.  But it must {\em not} 
be orthogonal to ${\bm{e}}_{p}$, along which the gradients are zero.  The 
components of the general stress-force equation for the rotating polytrope 
are, therefore, similar to (\ref{oned_poly_p_t})-(\ref{oned_poly_p_f}) and 
(\ref{oned_poly_w_t})-(\ref{oned_poly_rootg})
\begin{eqnarray}
T^{i}_{p} & = & 0 
\\
F_{p} & = & e^{i}_{r} [ p_{, i} \, + \, \rho \, (n+1) \, w_{, i} ] - \rho R E_{rot} \, e^{R}_{r} 
~~~~~~~~~~
\\
T^{i}_{w} & = & w_{, i} 
\\
F_{w} & = & \rho
\\
\sqrt{-g} & = & R
\end{eqnarray}
with a sum on $i$ over $R$ and $Z$, 
boundary conditions $p = 1$ and $e^{i}_{r} w_{, i} = 0$
at the stellar center and at the stellar surface
\begin{eqnarray}
f_{p} \; = \; p  \, - \, p_{s}
\\
f_{w} \; = \; w  \, - \, w_{s}
\end{eqnarray}
where $p_{s} << 1$ is a small fraction of the central pressure and $w_{s}$ is 
the specified surface potential.

Two different methods were tested for calculating $w_{s}$.  The first was an exterior  
multipole expansion 
\begin{equation}
\label {multipole_bc}
w_{s}^{M} \; = \; \sum_{\ell = 0}^{L} M_{\ell} \, r^{-\ell-1} \, P_{\ell} (Z/r) 
\end{equation}
where the $P_{\ell}$ are Legendre polynomials and $M_{\ell}$ are the
moments of the mass distribution in the star
\begin{equation}
M_{\ell} \; = \; \int_{\Omega} \rho (R,Z) \, r^{\ell} \, R \, dR \, dZ
\end{equation}
which, because of additional equatorial plane symmetry, are non-zero only
for even $\ell$.  Generally, orders up to $L=12$ were used. 
The second method used a Green's function integral over the domain
\begin{eqnarray}
w_{s}^{I} & = & \int_{\Omega} \frac{\rho}{|\bm{x} - \bm{x_{s}}|} \, d \Omega \nonumber
\\
& = & \int_{\Omega} \frac{\rho(R,Z) \, R \, dR \, dZ} {[R^2 + R_{s}^2 + (Z-Z_{s})^2]^{1/2}} \nonumber
\\
& & \times \; I_{e} \left(\ln \left[ \frac{(R+R_{s})^2 + (Z-Z_{s})^2} 
{(R-R_{s})^2 + (Z-Z_{s})^2} \right] ^{1/2} \right)
~~~~~
\label {integral_bc}
\end{eqnarray}
This expression is valid for all continuous, Newtonian self-gravitating, axisymmetric 
systems.\footnote{The integral in equation (\ref{integral_bc}) is over 
both northern and southern hemispheres of the star.  If, as is often the case, reflection symmetry 
is assumed at the equator and $\Omega$ refers only to the northern hemisphere, then the 
integral must take into account the southern contribution as well by summing one term as above 
plus one term with $(Z-Z_{s})$ replaced by $(Z+Z_{s})$.}  
The single-parameter complete elliptic integral
\begin{equation}
I_{e}(\alpha) \; \equiv \; \int_{0}^1 \, \frac{ds} {(1 - \tanh \alpha \, \cos 2 \pi s)^{1/2}}
\label {ellipt_int}
\end{equation}
represents the summed relative contributions to the surface potential at $(R_{s}, Z_{s})$ from 
different angular elements of a ring of matter at $(R, Z)$.  $I_{e}$ diverges with $\alpha$ 
(a measure of how close the ring is to $\bm{x_s}$), 
but it can be evaluated numerically easily and tabulated to a part in $10^{10}$ accuracy for the 
useful range $0 \leq \alpha \leq 10$ ({\it i.e.}, for rings approaching within only a fraction 
$2 \times e^{-10} = 9.1 \times 10^{-5}$ of $|\bm{x_{s}}|$) over which the integral lies in the range 
$1.0 \leq I_{e} \leq 5.1$.  In all models presented here, even with the largest meshes ($33^2$)
{\em and} adaptive gridding ($R_{s}/\delta R_{s} \gtrsim 200$), $\alpha$ remains well below 8.0 
and $I_{e}$ below $4.2$.

As implemented in the author's code, the speed of the integral outer boundary condition technique 
was significantly slower than the exterior multipole expansion, increasing 
the time to form the stiffness matrix (although not affecting the time to invert it) by 
factors of several.  While the author made no attempt to optimize the implementation, 
even after such efforts it nevertheless should remain somewhat expensive, as it requires 
(for each non-zero stiffness matrix element and each right-hand-side vector element) the 
complete integration of the potential (equation \ref{integral_bc}) at 4-6 surface points, 
each integration being the equivalent of of a multipole moment computation.  
However, while the multipole expansion began to break down for modest rotation speeds, the 
integral technique converged with no problem for all rotation speeds up to breakup, making 
the extra computational effort worthwhile and necessary.  In addition, as the integrals are 
done only for surface points, this hybrid technique (differential equation with integral 
boundary conditions) still will be much cheaper than computing the global potential by 
performing such an integral for every point in the {\em domain}.

Figure \ref{2dsph_poly9_fig} shows two dimensional $n=1$ non-rotating polytropes for the two element 
classes each with $9^2$ nodes --- the analog of the middle and right panels of 
Figure \ref{1dsph_poly_fig} --- using the multipole boundary condition.  
Note especially the variable grid spacing near the stellar surface and the 
difference in smoothness between linear and quadratic interpolation.
Figure \ref{2dsph_poly33_fig} shows the same models for the $33^2$ node cases.  Note especially
the departure from sphericity in the pressure contours in the models
employing linear elements that is absent in the quadratic element cases; 
the $9^2$ quadratic model is more accurate than the $33^2$ linear model.
However, errors in the two cases scale only as 
\begin{eqnarray*}
{\sf{E}}^{L2}_{linear} & \propto & (\aleph \wp)^{-1.0}
\\
{\sf{E}}^{L2}_{quadratic} & \propto & (\aleph \wp)^{-2.0}
\end{eqnarray*}
which is one full order less accurate than expected.  This may be due
to the {\em reflective} boundary conditions along the axis and equator, which are
only first and second order accurate, respectively, in the linear and
quadratic cases.  The nature of this boundary condition cannot be improved 
at this time, but will be once iterative/multigrid methods for solving the
coupled equations are implemented.  This will allow boundary ``ghost'' 
elements to be handled easily, allowing application of boundary conditions
as accurate as the mesh interior.

The Maclaurin spheroid sequence (a series of $n=0$, uniform density polytropes)
provides a full two-dimensional test of the method.  (Note that the method
takes no advantage of the uniform rotation, uniform density, or
polytropicity, so the ability of the code to solve the Maclaurin problem is a
good indication of how it will do on general multidimensional stellar 
structure and other more complex problems.)  
The appearance of Maclaurin spheroids is similar to that of the $n=1$ polytropes in 
Figure \ref{2dsph_poly9_fig} and 6, but the logarithm of the pressure varies little with radius 
except near the surface, where the radial grid spacing decreases dramatically due to 
the sudden pressure drop.  For this reason it is sufficient to use a larger pressure 
boundary value ($p_{s} = 10^{-2}$ rather than $10^{-4}$) in order to determine 
the locus of the Maclaurin spheroid surface.  
Figure \ref{2drot_poly33_fig} shows the analogy of 
Figure \ref{2dsph_poly33_fig} for a Maclaurin spheroid with ${\omega}^2/2 \pi G \rho = 0.224$
--- very near the theoretical limit of $0.2246656$ (\cite{tassoul78}).  The integral 
boundary condition (\ref{integral_bc}) is used to compute the surface potential.

The complete Maclaurin series from ${\omega}^2/2 \pi G \rho = 0$ to $0.224$ 
was computed in four different ways, using 
$9^2$ and $17^2$ nodes for the two classes of elements (linear and quadratic).  
Values of $\tau \equiv E_{rot}/|E_{w}|$, the rotational flattening ratio of the
semi-minor and semi-major axes $f = 1 - a_Z/a_R$, and total angular momentum 
$J = E_{rot}^{\frac{1}{2}} \int_{\Omega} \rho (R,Z) R^2 \, d\Omega$ have been 
computed from these models and are compared in Figure \ref{2drot_poly17_plots} with 
analytic curves from \cite{tassoul78} and \cite{chandra69}.  The fractional errors
for the four series are shown in Figure \ref{2drot_poly17_errs}.  The models are 
quite accurate for such a small number of elements, with errors in the $10^{-3}$ 
to $10^{-4}$ range in the $17^2$ quadratic case.  They show the expected result 
that third order interpolation is significantly more
accurate than second order, but the decrease in the errors with increasing
numbers of elements is not as steep as expected.  Most of this behavior is 
probably due to the less accurate reflective boundary conditions mentioned 
earlier.

\section{Discussion}

\subsection{Summary}

This paper has developed a general method for solving multidimensional structural, 
and dynamical, problems of astrophysics.  Virtually all situations involving 
continuous media are potentially addressable --- in normal flat Cartesian space 
or in curved spacetime.  Problems in this area include, but are not limited to, 
the full structure and secular evolution of viscous, rotating (and even magnetized) 
stars and accretion disks in two and three dimensions, interacting binaries, 
asymmetric stellar envelopes and winds, non-radial pulsating stars, nonlinear 
development of secular and thermal accretion disk instabilities, and stationary 
or evolving spacetimes.  

While this method is most useful for structures that evolve on timescales long 
compared to a dynamical time, there is no formal restriction on how short the 
evolution time must be.  
Therefore, the approach to dynamical instabilities from a stable configuration, 
and even initial dynamical development, also can be studied, although the 
author still recommends the use of an explicit code for full dynamical 
evolution.  

The equations of continuum astrophysics have been condensed into a general 
compact covariant form, and that form encoded into the author's FEM program 
GENRAL. A user can solve a particular astrophysics problem by supplying a 
single subroutine that takes as input one given coordinate position, plus the 
value of the variables and their gradients there, and returns as output the 
{\em differential} equations ({\it i.e.}, the generalized ``stress tensor'', 
``body force vector'', and possible boundary conditions appropriate for that 
problem).
The program then generates the nodal or 
``difference'' equations on a user-specified general curvilinear grid using the 
finite element weighted residual integrals, and solves the large set of coupled 
equations to produce the solution to the equations.  While described by discrete 
nodal values, as in finite difference methods, the finite element solution is continuous, 
as in spectral methods, due to the finite element interpolation functions.  
Either second order (linear interpolation) or third order (quadratic interpolation) 
accurate solutions are possible in the code.  In addition, the positions of the nodes 
themselves can be part of the solution (a ``rubber'' mesh), allowing grids to be 
fit to unknown boundary shapes and regions of high gradients to be more finely 
resolved with the same number of mesh points.

While the method is cast in a full covariant form, it is anticipated that the initial 
applications will be mainly in the area of non-relativistic stars or accretion disks 
in static gravitational fields.  The covariant form, however, is important even for 
non-relativistic problems.  When the mesh extends into the time domain, even only 
for one or two elements, the coordinate transformations that are a natural component of 
the finite element method {\em automatically} generate any arbitrary Lagrangian-Eulerian 
(ALE) advective derivatives needed to take possible grid motion into account.

The method has been demonstrated on astrophysically interesting problems (spherical or 
rotating polytropic stars) in one and two dimensions, with full adaptive gridding, 
and on simpler problems in three and four dimensions.

\subsection{A Note on the Solution of Elliptic Potential Problems}

A great deal has been written on the numerical solution of astrophysical potential 
problems like equation (\ref{poissons_eq}).  The technique used here has elements 
of past approaches plus some new features and is well-suited to the FEM.  
Like many past authors (\cite{clement78}; \cite{bgm98}), the approach 
here casts the problem as a differential equation with boundary conditions specified 
on the exterior of the domain.  However, rather than being a simple $1/r$ 
or low-order multipole potential at a large radius in the 
vacuum region, the author's preferred boundary condition is an integral solution of the 
differential equation, specified at the stellar surface. This integral is physically 
equivalent (on that surface) to the ``full integral'' technique that computes the potential 
{\em throughout the computational domain} using a Green's function integral rather 
than solving the differential equation itself (\cite{tassoul78}; \cite{em85}; \cite{keh89}).  
However, the author's method of evaluating this integral is somewhat different as it 
requires no expansion in terms of Legendre polynomials, relying instead on a single, 
slowly-varying, numerically-tabulated function $I_{e}(\alpha)$ to handle the axisymmetry 
of the potential field.  When calculated in this 
manner, using the integration techniques already available in the FEM, the numerical 
integral is a solution of the discrete finite element equations themselves to within the 
truncation error.  The boundary condition and interior differential equation, therefore, 
match well, leading to good convergence of the models.  

This technique will work in any situation where the full integral technique can be 
used:  extension of the Newtonian case into three dimensions will be trivial, and it will 
be straightforward for the general relativistic case as well.  In three dimensions there 
is no axisymmetry, so $I_{e}(\alpha)$ will not be needed, and $|\bm{x} - \bm{x_{s}}|$ will 
take the simple Pythagorean form.  For axisymmetric relativistic stars, the four metric 
potentials are given by three Green's function integrals plus a first order equation 
(\cite{keh89}).  Therefore, the three elliptic equations can be solved in the same manner 
as Poisson's equation is solved here (although probably using different tabulated 
$I_{e}$ functions), preserving the differential equations in the stellar interior but 
using the integral solution on the surface.  The fourth equation would be handled with 
the Petrov-Galerkin scheme demonstrated in section 4.  The advantage of this approach 
compared to the full integral technique is speed (the number of volume integrals is proportional 
to the domain surface area, not the volume).  The advantage over the multi-domain technique 
is convenience (one does not have to deal with the vacuum region and the 
matching of stellar and vacuum solutions).

\subsection{Unresolved Issues}

While the code and method are mature enough to begin solving two-dimensional structural 
problems routinely, there are several unresolved issues, mentioned in the text, that 
must be addressed more completely before the full potential of the astrophysical 
finite element method can be realized.

First and foremost are the execution speed and memory issues.  While the reader 
may consider the generation of the transformations and finite element integrals 
rather time-consuming, by far the greatest use of computer resources is the 
technique currently used to solve the coupled equations --- the Henyey 
technique.  For large meshes in three or more dimensions, it becomes 
prohibitively expensive, requiring thousands to many millions of {\em years} of 
CPU time ($\propto [\aleph \wp]^{3D-2}$) and equally absurd amounts of memory 
($\propto [\aleph \wp]^{2D-1}$) to invert once.  However, multigrid 
methods (\cite{brandt77}) need only about twice the grid size in storage 
($\propto [\aleph \wp]^D$) and only require a few sweeps of the mesh to 
converge ($\propto [\aleph \wp]^D \log [\aleph \wp]^D$).  
The author and P. Godon have been experimenting with modern parallel multigrid 
algorithms in finite {\em difference} codes with considerable success.  
The CPU-time and memory scalings, and linear speedup on parallel 
supercomputers, all have been realized.  Efforts are currently underway to 
make GENRAL a parallel, multigrid FEM code.  

Implementation of iterative schemes like multigrid for solving the equations 
will make straightforward the application of accurate reflective and periodic 
boundary conditions. 
While possible with the Henyey technique, this process is much more difficult 
as it involves columns of matrix elements far from the diagonal and special 
techniques for inversion.  With iterative methods, as with explicit codes, one 
can enclose the computational domain in a layer of ``ghost'' elements whose 
properties are determined at each iteration by the interior solution. 
The ghost element approach will have the same order of accuracy as the 
interpolation scheme, unlike the current approach for the reflective boundary 
condition, which uses essentially a backward difference.  

Another possibly important issue is time evolution.  All examples in this 
paper, even the four-dimensional ones, are time-independent and use a 
Cartesian metric.  The inclusion of time dependence may be as simple as 
employing a Minkowski metric and time derivatives of the variables, and 
the letting the finite element machinery solve the problem.  Often, 
however, the addition of a new feature generates new numerical problems 
which require modification of that machinery.  Until more experience is 
obtained with time dependent problems, it is not clear whether the 
techniques discussed here are complete or whether they will need additional 
major development to handle evolutionary situations.

Finally, many issues remain in the use of the finite element method for 
{\em dynamical} evolution problems.  These are currently important topics in the 
engineering field, but, because explicit finite difference codes do well for 
astrophysical problems in this area, development of these issues here will 
have lower priority.  They include adaptive mesh refinement (for 
dynamical collapse situations), implementation of the general boundary 
conditions in equation (\ref{general_bc}) (for magnetohydrodynamics and solving 
Maxwell's or Einstein's equations), and proper upwinding schemes with behavior 
comparable to the higher order Godonov schemes (for problems that develop 
shocks).

\subsection{A Note on Numerical Relativity with Finite Element Analysis}

In numerical relativity it is customary to perform a ``3+1 split'' of the 
metric such that
\begin{equation}
ds^2 \; = \; (- \alpha^2 \, + \, \beta_i \beta^i) dt^2 \, + \, 
2 \beta_i \, dx^i \, dt \, + \gamma_{ij} \, dx^i \, dx^j
\end{equation}
where $\gamma_{ij}$ is the 3-metric that raises or lowers indices on the shift 
3-vector $\beta_i$, and $\alpha$ is the ``lapse function'', all of which are 
functions of position in spacetime (\cite{adm}; \cite{york79}).  
The 3-metric is specified on the initial hypersurface by solving the field 
constraints (initial value data), the lapse and shift are computed from 
four coordinate (or ``gauge'') conditions, and the Einstein field equations 
are used to evolve $\gamma_{ij}$ to the next hypersurface.  
The goal is to choose a gauge in which the hypersurfaces do not intersect 
a singularity before a significant amount of evolution occurs in some part of 
the mesh.  The current method for singularity avoidance is to 
eliminate pathological parts of spacetime from the mesh (``excise the black
hole'') (\cite{cook98}).  

While such an approach is also possible with the FEM (\cite{amp98}, advancing 
time in a step-by-step fashion, the full covariant nature of the method and 
the lifting of the degeneracy between basic and mesh coordinate systems, allow 
additional approaches to be taken.  In particular, it becomes possible to 
extend the mesh fully in the time dimension, from initial to final 
hypersurface, choosing a relatively simple gauge for $\alpha$ and $\beta_{i}$.  
Then, adaptive gridding in all four dimensions can be used to keep the grid 
boundaries away from singularities and to further adjust the separation in 
time between spacelike hypersurfaces.  
Because of the ispoarametric transformation, the foliation no longer has to 
be along surfaces of constant time $x^0$.  
The separation between adjacent surfaces can be non-uniform, the time 
coordinate can vary considerably over the hypersurfaces, and the 
final hypersurface even can end at different times.  
In effect, the adaptive gridding completes, in mesh coordinates, the job 
of slicing and singularity-avoiding that a poor gauge choice may fail to do.
One advantage of this approach is that some or all of the field constraints 
can be applied on the final, instead of initial, hypersurface, turning 
an explicit hyperbolic problem into an implicit boundary value problem (like 
stellar structure) and possibly stabilizing the growth of errors.

However, while such techniques probably can succeed in keeping physical 
singularities at bay, it is doubtful that they can avoid coordinate 
singularities in general situations.  
(These arise in the most benign of curved surfaces --- 
on the surface of the earth, for example.)  Apart from knowing the geometry 
before hand and choosing the proper basic coordinate system, there are only a 
few ways to avoid these problems entirely.  
One is to embed the spacetime in a higher dimensional, flat Minkowskian space.  
In principle, as many dimensions as independent spacetime metric coefficients 
(ten) would be needed for the embedding, although it may be possible with fewer. 
From the ten hyperspace coordinates, and how they vary in the four-dimensional 
mesh, one then could derive the local metric of the spacetime and use 
it in the physical equations.  
These ten equations for the metric in terms of the hyper-coordinates, plus 
the six Einstein equations and the four adaptive gridding equations, would be 
sufficient to determine the twenty independent $g_{\alpha \beta}$ and 
hyper-coordinates at each node in the finite element mesh.  
While a fairly immense job for present-day computers, this prescription has 
the advantage of being singularity-free in general situations.  

Another method of avoiding coordinate singularities using the FEM is to 
dispense with global coordinates entirely, using only line segment lengths 
and deficit angles to describe the geometry and the Regge calculus to describe 
the physics (\cite{regge61}; \cite{holst98}).  At present, this approach has 
been developed only for simplex-type elements and not hypercubes, so it is not 
a straightforward application of the code discussed herein.  However, it may  
be useful to recast the Regge calculus for other element types.  

Finally, all methods that involve a full four-dimen-sional finite element spacetime 
are probably well beyond the capabilities of present computer technology, even 
with the use of parallel multigrid techniques.  Nevertheless, they appear to have 
such attractive features and elegance that it is important to begin to develop  
them.

\acknowledgments

The author is grateful to J. Fanselow for support during the early development of 
this work, to L. Caroff and M. Bicay at NASA for allowing a small portion of a 
theoretical astrophysics grant to be used for this purpose, and to the JPL Director's 
Research and Development Fund for support to complete this work.  Discussions 
with numerous people were very helpful, including G. Lyzenga and A. Raefsky on the 
finite element method, M. Norman on adaptive gridding, P. Godon on spectral methods 
and multigrid methods, and W. Cannon, S. Finn, M. Holst, 
K. Thorne, and J. York on the use of these techniques for general relativity.
This research was carried out at the Jet Propulsion Laboratory, California 
Institute of Technology, under contract to the National Aeronautics and Space 
Administration.



\begin{appendix}{}
\section{Appendix A.  Casting of the Differential Equations of Continuum Astrophysics into General 
\label{diffeq_app}
Finite Element Form}

This appendix shows that virtually all the equations of astrophysics of continuous media can be 
cast into the flux-conservative, finite element form (equation \ref{general_eq}) and 
their boundary conditions into equation (\ref{general_bc}).  That is, while possibly 
second order in spatial and time derivatives, they can be written as the four-divergence 
of a generalized stress tensor plus a generalized body force vector, each of 
which are functions of no more than the first spacetime derivative (four-gradient) of 
the variables.  
Of course, it is always possible to define additional 
variables ({\it e.g.}, the 24 connection coefficients) and turn the field 
equations and conservation laws into first-order equations involving only the $F_{q}$ 
term in equation (\ref{general_eq}).  The challenge, however, is to use only the original 
metric and field components as variables (avoiding additional computational expense), 
and still maintain the flux-conservative form.  Below is one solution to this problem.

\subsection{The Equations in Geometric Form}

The discussion here is concerned only with differential equations.  
Local physics, such as the equations of state, opacity, emissivity, 
viscosity, {\it etc.}, is not treated in detail.  While having position and time 
dependence, these processes can be described with simple algebraic equations 
that do not affect the numerical method used.  

The differential equations are the deceptively simple 
set of the Einstein equations for the gravitational field
\begin{equation}
\label {einstein_eqs}
\bm{\cal{G}} \; = \; 8 \pi \, \bm{\cal{T}}
\end{equation}
(where $\bm{\cal{G}}$ and $\bm{\cal{T}}$ are the symmetric Einstein curvature and 
stress-energy-momentum [SEM] tensors), and Maxwell's equations for the electromagnetic 
field
\begin{eqnarray}
\label {maxwell1}
\bm{\nabla \cdot \, \cal{F}} & = & 4 \pi \, \bm{\cal{J}}
\\
\label {maxwell2}
{\bm{\nabla \cdot \, \cal{M}}} & = & 0
\end{eqnarray}
where ${\bm{\nabla}}$ is the covariant gradient operator, the antisymmetric Maxwell 
tensor ${\bm{\cal{M}}} = {\bm{^{\ast} \cal{F}}}$ is the dual of the antisymmetric 
Faraday tensor ${\bm{\cal{F}}}$, and ${\bm{\cal{J}}}$ is the four-current.  
With the symmetry, equations (\ref{einstein_eqs}) are 10 in number, and 
(\ref{maxwell1})-(\ref{maxwell2}) constitute 8, for a total of 18.  However, because 
of identities satisfied by the Einstein and Faraday tensors, there are actually only 12 
{\em independent} equations (6 metric and 6 electromagnetic) but 16 unknowns at each 
point in space:  the 10 independent components of the metric ${\bm{g}}$ and 
the 6 independent components (the electric and magnetic fields ${\bm{E}}$ and 
${\bm{B}}$) of the antisymmetric Faraday tensor.  
The remaining 4 metric unknowns are determined by the choice of a coordinate 
system or gauge.

The standard method for generating a set of 12 evolutionary equations is to 
project (\ref{einstein_eqs})-(\ref{maxwell2}) into the hypersurface normal to 
a time-like vector (or world line) ${\bm{n}}$ with the projection tensor 
(equation \ref{projection_tensor}).  
For example, if ${n}_{\mu} = \, g_{0 \mu}/\sqrt{-g_{00}})$, then only the 
spatial part ${\cal{S}}_{ij}$ will be non-zero, with $i,j=1,2,3$.  In general, 
however, ${\bm{n}}$ can be any time-like vector, so the equations will be left in 
general form.  If the twelve {\em spacelike} components of equations 
(\ref{einstein_eqs})-(\ref{maxwell2}) are satisfied throughout the 
four-dimensional spacetime domain $\Omega$ (with one factor of ${\bm{\cal{S}}}$ 
for each tensor order)
\begin{eqnarray}
\label {field_eq1}
{ \bm{\cal{S} \cdot {\cal{G}} \cdot {\cal{S}}} } & = 
& 8 \pi \, { \bm{\cal{S} \cdot {\cal{T}} \cdot {\cal{S}}} } 
\\
\label {field_eq2}
{ \bm{\cal{S} \cdot }} \, ( { \bm{\nabla \cdot \cal{F}} } ) & = 
& 4 \pi \, { \bm{\cal{S} \cdot \cal{J}} } 
\\
\label {field_eq3}
{ \bm{\cal{S} \cdot }} \, ( { \bm{\nabla \cdot \cal{M}} } ) & = & 0
\end{eqnarray}
then all that is necessary to satisfy the {\em timelike} components 
\begin{eqnarray}
\label {constr1}
{\bm{n \cdot \cal{G}}} & = & 8 \pi \, {\bm{n \cdot \cal{T}}} 
\\
\label {constr2}
{\bm{n \cdot \nabla \cdot \cal{F}}} & = & 4 \pi \, {\bm{n \cdot \cal{J}}} 
\\
\label {constr3}
{\bm{n \cdot \nabla \cdot \cal{M}}} & = & 0
\end{eqnarray}
throughout all spacetime is to satisfy the latter equations 
on {\em one} hypersurface only. 
Equations (\ref{field_eq1})-(\ref{field_eq3}), therefore, are the 12 
independent differential equations to be solved for the six metric and six 
electromagnetic field components, while equations 
(\ref{constr1})-(\ref{constr3}) are the {\em constraints} that need to be 
satisfied in order for a solution to exist at all.
(For reference, equation (\ref{field_eq2}) is Ampere's law, (\ref{field_eq3}) 
Faraday's law, (\ref{constr1}) contains the Hamiltonian and momentum 
constraints [by further contraction with ${\bm{n}}$ or ${\bm{\cal{S}}}$, 
respectively], (\ref{constr2}) is Coulomb's law, and (\ref{constr3}) is the 
solenoidal condition on the magnetic field.)

Equations (\ref{field_eq1}), with (\ref{constr1}) as initial conditions, 
constitute the Cauchy problem of general relativity.  Equations 
(\ref{field_eq3}) and (\ref{constr3}) are the covariant form of the 
Evans-Hawley ``constrained transport'' method for enforcing the solenoidal 
constraint in {\em non-relativistic} magnetohydrodynamics (\cite{eh88}).
Equations (\ref{field_eq2}) and (\ref{constr2}) represent constrained 
transport in the presence of sources.  When (\ref{field_eq1})-(\ref{field_eq3}) 
are solved as Cauchy problems, equations (\ref{constr1})-(\ref{constr3}) are 
applied on the {\em initial} hypersurface.  However, as our approach here is 
to relax the system for a {\em four}-dimensional spacetime, instead of 
evolving a three-dimensional surface forward in time, they can be applied on 
{\em any} spacelike hypersurface.  

In addition to the field equations, there are conservation laws that 
follow from identities satisfied by the fields.  The Einstein curvature tensor 
is constructed in such a way that ${\bm{\nabla \cdot \cal{G}}} = 0$, so the 
conservation of energy and momentum
\begin{equation}
\label{cons_laws}
{\bm{\nabla \cdot \cal{T}}} = 0
\end{equation}
must also hold from equation (\ref{einstein_eqs}).  Similarly, as 
${\bm{\cal{F}}}$ satisfies 
${\bm{\nabla \cdot}} \, ({\bm{\nabla \cdot {\cal{F}}}}) \; = \; 0$, then the 
four-current must also be conserved
\begin{equation}
\label{cons_charge}
\bm{\nabla \cdot \cal{J}} \; = \; 0
\end{equation}
The field equations then are ``closed'' by expressing the SEM tensor and 
four-current in terms of the state variables, and solving the 
conservation laws of energy and momentum for those variables. 
For most conceivable astrophysical situations --- including those with multi-fluid 
dynamics, electromagnetic fields and currents, radiation, viscosity, and 
nuclear reactions --- expressions for ${\bm{\cal{T}}}$ and ${\bm{\cal{J}}}$ 
involve terms with, at most, first-order derivatives of the state variables 
with respect to space or time.
This is true even in situations near black hole horizons 
where particle interaction and fluid flow time scales are comparable, and 
equations like Ohm's law, for example, are no longer valid.  

A final group of differential equations comes from forming the zeroth,
first, and second moments of the Boltzmann-Vlasov equation for each 
particle species {\it {\scriptsize{(j)}}} (photons, nuclei, {\it etc.}).  
These determine each species' individual number density, velocity (including 
the peculiar drift velocity ${\bs{q}}^{(j)}$), and internal energy per particle.
The kinetic equations give rise to familiar processes like nuclear burning, 
radiative transport, viscosity, and electrical conductivity.  Nevertheless, 
they all have the same ``conservative'' form, with the divergence of a term 
that involves (at most) first-order spatial derivatives of the state variables. 
For example, the zeroth moment of these kinetic equations yields 
\begin{equation}
\label {cons_particle}
\bm{\nabla \, \cdot} \, N^{(j)} \, [ {\bs{u}} \, + \, {\bs{q}}^{(j)} ] = c^{(j)}
\end{equation}
where $N^{(j)}$ s the particle number density and $c^{(j)}$ is the net particle 
creation rate due to nuclear reactions.  
For many astrophysical situations, particle conservation will be the only one 
of these equations needed, the drift velocity and particle energies being 
determined by the diffusion or other approximations.

\subsection{The Equations in Component Form}

With no loss of generality one can choose to write the differential equations 
in a coordinate frame.  In this case, the connection coefficients are given by 
\begin{equation}
\label {conn_coeff_def}
{\Gamma^{\alpha}}_{\beta \gamma} \; = \; \frac{1}{2} \, g^{\alpha \mu} \, 
(g_{\mu \beta , \gamma} \, + \, g_{\mu \gamma , \beta} \, - \, g_{\beta \gamma , \mu})
\end{equation}
and their trace reduces to 
\begin{equation}
\label {conn_coeff_trace}
{\Gamma^{\mu}}_{\beta \mu} \; = \; {(\ln \sqrt{-g})}_{, \beta} 
\end{equation}
where $g$ is the determinant of the metric
\begin{equation}
g \; \equiv \; {\rm det} || {\bm{g}} ||
\end{equation}
This yields simple expressions for the gradients and divergences in curved 
space, and the field equations (\ref{field_eq1}, \ref{field_eq2}, and 
\ref{field_eq3}) and conservation laws (\ref{cons_laws}, \ref{cons_charge}, 
and \ref{cons_particle}) 
become 
\begin{eqnarray}
\lefteqn{
[ \, {{\cal{S}^{\alpha}}_{i}} ( {\Gamma^{\mu}}_{\alpha \beta} {{\cal{S}^{\beta}}_{j}} \, - \, 
                             {{\cal{S}^{\mu}}_{j}} {\Gamma^{\beta}}_{\alpha \beta} ) 
\sqrt{-g} \, ]_{, \mu} \, - \, 
} 
~~~~~~~~~~~~~~~~~~~~~~~~~~~~~~~~~~~~~~~~~~~~~~~~~~ \nonumber 
\\ 
\lefteqn{~~
\{ \, 
{{\cal{S}^{\alpha}}_{i}} {{\cal{S}^{\beta}}_{j}} [ \, 
{{\Gamma^{\mu}}_{\alpha \nu}} {{\Gamma^{\nu}}_{\beta \mu}} \, - \, 
{{\Gamma^{\mu}}_{\alpha \mu}} {{\Gamma^{\nu}}_{\beta \nu}} \, ] \, + \, 
} 
~~~~~~~~~~~~~~~~~~~~~~~~~~~~~~~~~~~~~~~~~~~~~~~~ \nonumber 
\\ 
\lefteqn{~~~
( {{\cal{S}^{\alpha}}_{i}} {{\cal{S}^{\beta}}_{j}} )_{, \mu} {{\Gamma^{\mu}}_{\alpha \beta}} \, - \, 
( {{\cal{S}^{\alpha}}_{i}} {{\cal{S}^{\beta}}_{j}} )_{, \beta} {{\Gamma^{\mu}}_{\alpha \mu}} \, + \, 
} 
~~~~~~~~~~~~~~~~~~~~~~~~~~~~~~~~~~~~~~~~~~~~~~~ \nonumber 
\\ 
\label {comp_einstein}
8 \pi ( {{\cal{S}^{\alpha}}_{i}} {\cal{T}}_{\alpha \beta} {{\cal{S}^{\beta}}_{j}} - \frac{1}{2} {\cal{S}}_{ij} {\cal{T}} ) \, 
\} \, \sqrt{-g} & = & 0
\\
\lefteqn{
( {\cal{S}}_{i \mu} {\cal{F}}^{\mu \beta} \, \sqrt{-g} )_{, \beta} \, - \,
} 
~~~~~~~~~~~~~~~~~~~~~~~~~~~~~~~~~~~~~~~~~~~~~~~~~~ \nonumber
\\
\label {comp_ampere}
\left [ \, {\cal{S}}_{i \mu , \beta} {\cal{F}}^{\mu \beta} \, + \, 
4 \pi {\cal{S}}_{i \mu} {\cal{J}}^{\mu} \, \right ] \, \sqrt{-g} & = & 0 
\\
\label {comp_faraday}
( {\cal{S}}_{i \mu} {\cal{M}}^{\mu \beta} \, \sqrt{-g} )_{, \beta} 
- {\cal{S}}_{i \mu , \beta} {\cal{M}}^{\mu \beta} \, \sqrt{-g} & = & 0
\\
\label {comp_sem_cons}
( {{\cal{T}}^{\alpha \beta}} \, \sqrt{-g} )_{, \beta} \, + \, 
{\Gamma^{\alpha}}_{\mu \beta} {\cal{T}}^{\mu \beta} \, \sqrt{-g} & = & 0
\\
\label {comp_chg_cons}
( {{\cal{J}}^{\beta}} \, \sqrt{-g} )_{, \beta} & = & 0
\\
\label {comp_continuity}
( N^{(j)} [ u^{\beta} \, + \, q^{(j) \beta}] \, \sqrt{-g} )_{, \beta} \, - 
\, c^{(j)} \, \sqrt{-g}  & = & 0 ~~~~~~~~~~
\end{eqnarray}
which are all in the conservative finite element form (\ref{general_eq}).
Use has been made of the expression for the Einstein tensor
\begin{displaymath}
{\cal{G}}_{\alpha \beta} \; = \; {\cal{R}}_{\alpha \beta} \, - 
\, \frac{1}{2} g_{\alpha \beta} {\cal{R}}
\end{displaymath}
in terms of the Ricci tensor
\begin{displaymath}
{\cal{R}}_{\alpha \beta} \; = \; \frac{1}{\sqrt{-g}} 
( {\Gamma^{\mu}}_{\alpha \beta} \, \sqrt{-g} )_{ , \mu} \, - 
\, ( \ln \sqrt{-g})_{, \alpha \beta} \, - 
\, {\Gamma^{\mu}}_{\nu \alpha} {\Gamma^{\nu}}_{\beta \mu}
\end{displaymath}
where the Ricci scalar is the contraction ${\cal{R}} \equiv {{\cal{R}}^{\mu}}_{\mu}$, 
and the SEM scalar is the contraction ${\cal{T}} \equiv {{\cal{T}}^{\mu}}_{\mu}$.  
Note that, as equations (\ref{comp_einstein}) 
are symmetric in $i$ and $j$, they are six in number, and (\ref{comp_ampere}) 
and (\ref{comp_faraday}) are three equations each.  The term inside the divergence 
in equation (\ref{comp_einstein}) is the trace-free curvature --- the ``flux'' of 
the metric; the ``force'' terms involve derivatives of the projection tensor and 
a matter source generating the curvature.  


Boundary conditions on the conservation laws are generally of the simple form 
(\ref{simple_bc}).  Furthermore, 
it is well known that the Einstein field constraints do not depend on second
derivatives with respect to time, and the Maxwell constraints do not depend on time 
derivatives at all.  So, it should be possible to write these in the form 
(\ref{general_bc}), with at most a divergence on the boundary and a gradient
normal to it.  One of the more compact versions of this is 
\begin{eqnarray}
\lefteqn{
( {\cal{S}}^{\alpha \beta} {\Gamma^{\mu}}_{\alpha \beta}  \sqrt{-g} )_{, \mu} \, - \, 
( {\Gamma^{\mu}}_{\alpha \beta}  \sqrt{-g} )_{, \beta} 
{\cal{S}}^{\alpha \beta} \, - \, 
} 
~~~~~~~~~~~~~~~~~~~~~~~~~~~~~~~~~~~~~~~~~~~~~~~~~~~~~ \nonumber 
\\ 
\lefteqn{~~~
[ \, {\cal{S}}^{\alpha \beta} ( {\Gamma^{\mu}}_{\alpha \nu} {\Gamma^{\nu}}_{\beta \mu} - 
{\Gamma^{\mu}}_{\alpha \mu} {\Gamma^{\nu}}_{\beta \nu} )  \, + \, 
{{\cal{S}}^{\alpha \beta}}_{, \mu} {\Gamma^{\mu}}_{\alpha \beta} \, + \, 
} 
~~~~~~~~~~~~~~~~~~~~~~~~~~~~~~~~~~~~~~~~~~~~~~~~~~ \nonumber 
\\ 
\label {comp_hamilton_constr}
8 \pi ( {n}^{\alpha} {n}^{\beta} {\cal{T}}_{\alpha \beta} - \frac{1}{2} {\cal{T}} ) \, 
] \, \sqrt{-g} & = & 0
\\ 
\lefteqn{
( {n}^{\beta} {{\cal{S}}^{\alpha}}_{i} {\Gamma^{\mu}}_{\alpha \beta} \sqrt{-g} )_{, \gamma} 
{{\cal{S}}^{\alpha}}_{\mu} - 
( {{\cal{S}}^{\beta}}_{\mu} {{\cal{S}}^{\alpha}}_{i} {\Gamma^{\mu}}_{\alpha \beta} \sqrt{-g} )_{, \gamma} 
{n}^{\gamma} \, - \, 
} 
~~~~~~~~~~~~~~~~~~~~~~~~~~~~~~~~~~~~~~~~~~~~~~~~~~~~~ \nonumber 
\\ 
\lefteqn{~~
\{ \, {{\cal{S}}^{\alpha}}_{i} [ {n}^{\beta} 
( {\Gamma^{\mu}}_{\alpha \nu} {\Gamma^{\mu}}_{\beta \nu} - 
  {\Gamma^{\mu}}_{\alpha \mu} {\Gamma^{\nu}}_{\beta \nu} \, + \, 
  8 \pi {\cal{T}}_{\alpha \beta} ) \, + \, 
} 
~~~~~~~~~~~~~~~~~~~~~~~~~~~~~~~~~~~~~~~~~~~~~~~~~~~ \nonumber 
\\ 
\lefteqn{~~~~
{n}^{\beta}         {\Gamma^{\mu}}_{\alpha \beta} {n}_{\mu , \gamma} {n}^{\gamma} \, + \, 
{n}^{\beta}_{, \mu} {\Gamma^{\mu}}_{\alpha \beta} \, ] \, + \, 
} 
~~~~~~~~~~~~~~~~~~~~~~~~~~~~~~~~~~~~~~~~~~~~~~~~~ \nonumber 
\\ 
\label {comp_momentum_constr}
{{{\cal{S}}^{\alpha}}_{i}}_{, \mu} ( {n}^{\beta} {\Gamma^{\mu}}_{\alpha \beta} \, - \,
{n}^{\mu} {\Gamma^{\nu}}_{\alpha \nu} \, ) \} \sqrt{-g} & = & 0
\\
\lefteqn{
( {n}_{\mu} {\cal{F}}^{\mu \nu} \, \sqrt{-g} )_{, \beta} {{\cal{S}}^{\beta}}_{\nu} \, - \,
} 
~~~~~~~~~~~~~~~~~~~~~~~~~~~~~~~~~~~~~~~~~~~~~~~~~~~~~ \nonumber 
\\
\label {comp_coulomb_constr}
\left [ \, {\cal{F}}^{\mu \nu} {n}_{\mu , \beta} {{\cal{S}}^{\beta}}_{\nu} \, - \, 
4 \pi {n}_{\mu} {\cal{J}}^{\mu} \, \right ] \, \sqrt{-g} & = & 0 
\\
\lefteqn{
( {n}_{\mu} {\cal{M}}^{\mu \nu} \, \sqrt{-g} )_{, \beta} {{\cal{S}}^{\beta}}_{\nu} \, - \,
} 
~~~~~~~~~~~~~~~~~~~~~~~~~~~~~~~~~~~~~~~~~~~~~~~~~~~~~ \nonumber 
\\
\label {comp_solenoid_constr}
{\cal{M}}^{\mu \nu} {n}_{\mu , \beta} {{\cal{S}}^{\beta}}_{\nu} \, \sqrt{-g} & = & 0
~~~~~~~~~
\end{eqnarray}
where $n^{\mu}$ is now identified as the boundary normal.  
Equations (\ref{comp_hamilton_constr}) and (\ref{comp_momentum_constr})
are, respectively, the Hamiltonian and momentum constraints.  

In addition to the Einstein field equations (\ref{comp_einstein}) four more 
coordinate conditions, and their possible boundary conditions, are needed to completely 
specify the metric.  As these
are rather arbitrary, these are left unspecified for now.  (Of
course, if they are to be differential equations, they also must be in the 
form (\ref{general_eq}).)

\setcounter{equation}{0}

\section{Appendix B.  Conditions on the Isoparametric Transformation}
\label{xform_cond_app}

Not all ${{\cal{L}}^{\alpha}}_{\alpha '}$ specified by equation (\ref{isoparm_general}) 
are valid transformations in all situations.  
Sometimes they must satisfy certain conditions, which are discussed below.  Two of these 
are conditions on the new (element mesh) metric formed by the transformation and the third is 
a condition on the element spacing itself.  However, only the first condition always needs 
to be satisfied for a stable mesh.  The other two are necessary only under certain 
circumstances.

\subsection{The Jacobi Condition}

The most important of these conditions is that the Jacobian determinant of the 
transformation be non-zero 
\begin{displaymath}
{\cal{L}} \; \equiv \; {\rm det} ||{\bm{\cal{L}}}|| \neq 0
\end{displaymath}
However, since 
${\cal{L}}^{2} g \, = \, g '$, and since $g < 0$, this also means that 
the determinant of the new metric must be negative
\begin{equation}
\label{jacobi_metric_cond}
g ' \; = \; {\rm det} ||{\bm{g '}}|| \; < \; 0
\end{equation}
The Jacobi condition (\ref{jacobi_metric_cond}) ensures that the unit vectors in the 
element (primed) system, while not necessarily orthogonal, nevertheless form an 
independent coordinate system that has an arrow of time.  
It is absolutely necessary that this condition be satisfied in order that the mesh 
be well behaved.

\subsection{The Local-Lorentz Condition}

A second possible condition is that the element coordinate system have a locally Lorentz 
character everywhere --- {\it i.e.}, that $\xi^{0'}$ be the mesh time coordinate.  
This ensures that, if the spatial portion of the element mesh moves with time, the mesh 
velocity always will be less than the speed of light.  This is important, however, 
only if the mesh is used as a frame of reference for measuring physical quantities.

There are various ways of ensuring the Lorentz nature of the transformation 
${{\cal{L}}^{\alpha}}_{\alpha '}$.  The safest and simplest way is to ensure that each unit 
vector in the new space satisfies the proper timelike or spacelike constraint.  With the 
constraints that $\xi^{0'}$ denotes the time dimension and that 
${\bm{e}}_{\alpha '} \cdot {\bm{e}}_{\beta '} = g_{\alpha ' \beta '}$, these local 
Lorentz conditions become
\begin{eqnarray}
\label {local_lorentz_timelike}
g_{0' 0'} & = & {{\cal{L}}^{\alpha}}_{0'} {{\cal{L}}^{\beta}}_{0'} g_{\alpha \beta} \; < \; 0
\\
\label {local_lorentz_spacelike}
g_{i' i'} & = & {{\cal{L}}^{\alpha}}_{i'} {{\cal{L}}^{\beta}}_{i'} g_{\alpha \beta} \; > \; 0
\end{eqnarray}
Inequality (\ref{local_lorentz_spacelike}) is equivalent to demanding that each of the element 
sides in the $\xi^{i'}$ direction be spacelike.

A less stringent, but still sufficient, condition on $g_{i' i'}$ could be derived by choosing 
a specific timelike vector, such as ${n}^{\mu '} = {\delta_{0'}}^{\mu '}/ \sqrt{-g_{0' 0'}}$ 
(which still requires $g_{0' 0'} < 0$), and then constructing from the corresponding projection 
tensor a set of three independent vectors orthogonal to ${n}^{\mu '}$
\begin{displaymath}
{s_{i'}}^{\mu '} \; = \; {n}_{i'} {n}^{\mu '} \; + \; {\delta_{i'}}^{\mu '}
\end{displaymath}
The condition that these vectors be spacelike (${\bm{s}}_{i'} \cdot {\bm{s}}_{i'} > 0$) leads 
to a modified form for inequality (\ref{local_lorentz_spacelike})
\begin{displaymath}
g_{i' i'} \; > \; -(g_{0' i'}/\sqrt{-g_{0' 0'}})^{2}
\end{displaymath}
In this less restrictive case, the ${\bm{e}}_{i'}$ unit vectors can be a bit timelike, but no more 
so than that given by the above inequality.  

\subsection{The Courant Condition}

In standard nonrelativistic computational fluid dynamics, in order that the (explicit) 
forward integration in time be stable, the distance traversed in a single time step by 
sound waves or by the fluid itself (whichever is faster) must be substantially 
less than the mesh spacing.  The ratio of these distances, called the Courant 
number ${\sf{C}}$, is chosen to be $\sim 0.1 - 0.4$ or so depending on the 
stability of the numerical integration scheme.  This nonrelativistic Courant 
condition (which can be written as $-\Delta t^2 / {\sf{C}}^2 v_{max}^2 \; + \; \Delta x^2 > 0$ 
for one-dimensional flow) easily generalizes in the four-dimensional general relativistic 
case to
\begin{displaymath}
g_{\alpha \beta} \, \Delta {\chi}^{\alpha} \, \Delta {\chi}^{\beta} \; > \; 0
\end{displaymath}
where the vector ${\bm{\Delta \chi}} \, =$ ($\Delta t/{\sf{C}} v_{max}, \Delta x, 
\Delta y, \Delta z$).  It further generalizes in the case of a general curvilinear 
element mesh to 
\begin{equation}
\label {courant}
g_{\alpha ' \beta '} \, \Delta {\zeta}^{\alpha '} \, \Delta {\zeta}^{\beta '} \; > \; 0
\end{equation}
in each element, with ${\bm{\Delta \zeta}} \, =$ ($\Delta \xi^{0'}/{\sf{C}} v'_{max}$,
$\Delta \xi^{1'}$, $\Delta \xi^{2'}$, $\Delta \xi^{3'}$ ); $\Delta \xi^{\alpha '}$ is 
the width of the element in each mesh dimension; and $v'_{max}$ is a three-velocity 
magnitude equal to the maximum disturbance speed within the element.  
Because stable implicit techniques are used in time as well as space, a 
Courant number very close to unity probably can be tolerated.  Therefore, in 
the case of relativistic flow, where $v'_{max} = 1$, the Courant 
condition reduces to the requirement that the geodesics connecting opposing 
corners in each element must be spacelike ($g_{\alpha ' \beta '} \, \Delta \xi^{\alpha '} \, 
\Delta \xi^{\beta '} \, = \, \Delta s^2 > 0$).  

Generally, the Courant condition (\ref{courant}) is much too restrictive and is 
routinely violated in slowly evolving or steady-state problems, where time steps 
are very long or even infinite.  In implicit codes the condition needs to be 
satisfied only if one wishes to follow every short timescale transient phenomenon or 
wave.

\end{appendix}

%
%

%



\begin{figure}{\bf{FIGURE CAPTIONS}}

\caption{
\label{shapes_fig} 
Examples of shape functions in one dimension for 
linear (top) and quadratic elements (bottom).  
The solid line shows the shape function for the $\hat{0}$ node 
(at $\xi = \xi_{\hat{0}} = 0$); the dashed line for the $\hat{1}$ node 
(at $\xi_{\hat{1}} = 0.25$); and so on.  
Note the element boundary nodes (large open 
circles) and interior nodes (smaller filled circles in the quadratic 
case).  The derivatives of the shape functions are discontinuous at boundary 
nodes, although the functions themselves are continuous.  Each function attains 
unit value at its corresponding node and exactly zero at all other nodes 
in the element.  By definition, shape functions are also identically zero in 
elements not containing their corresponding node.}

\caption{
\label{1dpoisson_fig} 
One-dimensional results for adaptive grid tests.  Top left: solution 
of the Fermi-Dirac function test with $f = 50$, 
using 8 linear, equally-spaced elements, and yielding a very large 
error (${\sf{E}}^{L2}_w = 0.27$); top right: using 16 equally-spaced elements 
(${\sf{E}}^{L2}_w = 0.027$); middle left: 8 adaptive elements 
(${\sf{E}}^{L2}_w = 0.031$); middle right: 16 adaptive elements 
(${\sf{E}}^{L2}_w = 0.0068$); bottom left: same as middle right, but plotted 
against $\log w$;  bottom right: results when solving for $\tilde{w} = 
\log_{10} w$.  The dotted line shows the exact solution.}

\caption{
\label{2dpoisson_fig}
Two-dimensional results for adaptive grid tests, showing contour 
plots of the solution $w$.  All use 
meshes of $16 \times 16$ linear elements with $f=50$, $a=2$, and 
$b=c=0$ in the Fermi-Dirac function.  In all cases, coordinates and differential 
equations are 
expressed in $x$ and $y$ only, and although derivatives are 
calculated on the curvilinear grid, they are immediately transformed to 
the $(x,y)$ system and used as such in the equations.
Top left: uniformly-spaced Cartesian grid (${\sf{E}}^{L2}_w = 0.081$);  
top right: circular-polar grid (${\sf{E}}^{L2}_w = 0.067$); middle left: elliptical grid
with the same axial ratio as the solution (${\sf{E}}^{L2}_w = 0.037$); 
middle right: adaptive elliptical grid allowing finer resolution of the Fermi 
surface (${\sf{E}}^{L2}_w = 0.010$); bottom left: same as middle right, but with logarithmic 
contours (note oscillations similar to those in the bottom left panel of Figure 2);  
bottom right: resulting solution when solving for $\tilde{w} = \log_{10} w$.}

\end{figure}

\begin{figure}

\caption{
\label{1dsph_poly_fig}
One-dimensional $n=1$ spherical polytropic stars with 9 nodes.  Left: standard 
Galerkin weighting of the FEM integrals, with linear elements (${\sf{E}}^{L2}_p = 1.4$);
middle: Petrov-Galerkin type 2 weighting and linear elements (${\sf{E}}^{L2}_p = 0.048$); 
right: same as middle, but with quadratic elements (${\sf{E}}^{L2}_p = 0.0018$).  
All models use adaptive gridding and logarithmic variables.  
Note the close nodal spacing near the stellar surface.}

\caption{
\label{2dsph_poly9_fig}
Two-dimensional $n=1$ spherical polytropic stars with $9^2$ nodes using 
the multipole boundary condition ($w_{s}^{M}$). 
Top panels: mesh and pressure contours for (bi-) linear elements;  bottom 
panels: mesh and pressure contours for (bi-) quadratic elements.  Contours follow 
the precise interpolation within the respective elements.  Note the adaptive gridding near 
the stellar surface, similar to that in Figure \ref{1dsph_poly_fig}.}

\caption{
\label{2dsph_poly33_fig}
Same as Figure \ref{2dsph_poly9_fig}, but with $33^2$ nodes.  Note the persistent
errors in the pressure contours for linear elements which are absent
in quadratic elements with far fewer nodes.}

\caption{
\label{2drot_poly33_fig}
Similar to Figure \ref{2dsph_poly33_fig}, but using the integral boundary condition 
($w_{s}^{I}$) and solving a two-dimensional $n=0$ rotating polytropic star 
(Maclaurin spheroid) with $\omega^2/2 \pi G \rho = 0.224$ --- very close 
to the theoretical limit of $0.2246656$.}

\caption{
\label{2drot_poly17_plots}
Comparison of rotating Maclaurin spheroid model stars, computed with
different finite element meshes, with the analytic solution:  
Top panels: linear elements with $9^2$ and $17^2$ nodes, respectively; 
bottom panels: quadratic elements with $9^2$ and $17^2$ nodes.  Although
$\omega^2$ is the primary model parameter, the
flattening ratio, total angular momentum, and $\omega^2$ itself, with the normalizations 
in Tassoul (1978), are plotted as a 
function of the derived parameter $\tau$ for comparison with the analytic theory.}

\caption{
\label{2drot_poly17_errs}
Absolute value of the fractional errors for the Maclaurin spheroids in 
Figure \ref{2drot_poly17_plots} as a function of $\omega^2$.  Solid lines 
show models with linear elements, dashed lines quadratic elements; curves 
without symbols are for $9^2$ node models, those with are for $17^2$ models.}

\end{figure}

%

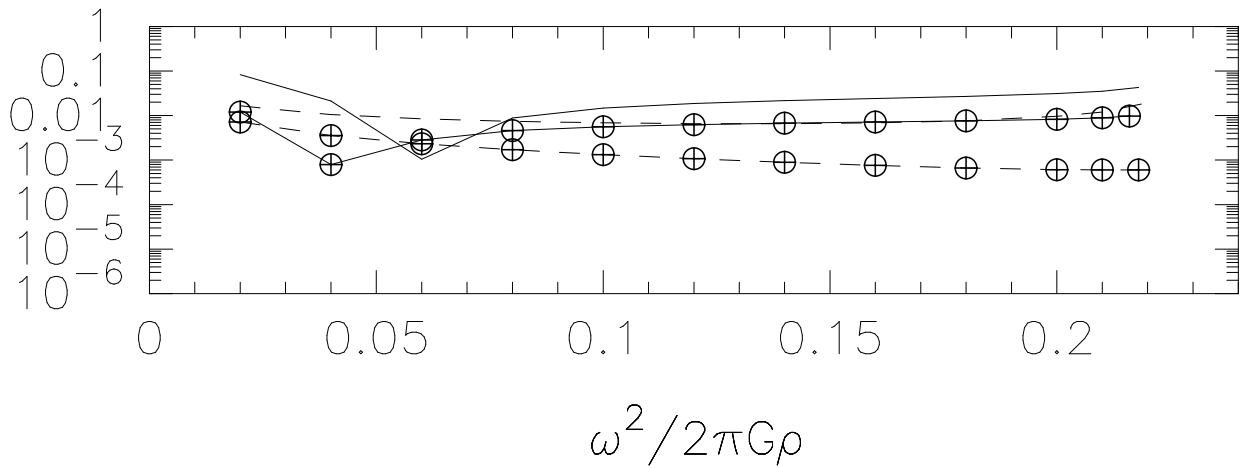
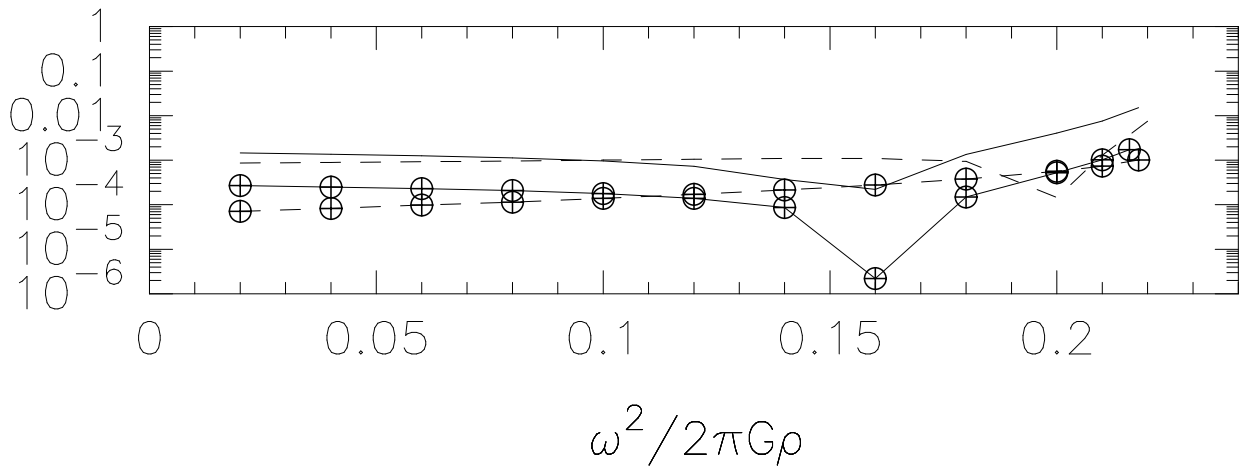
\end{document}